\documentclass[letterpaper,twocolumn,10pt]{article}

\usepackage{tikz}
\usepackage{amsmath}

\usepackage{filecontents}

\usepackage{tikz}
\usetikzlibrary{shapes,arrows}
\usetikzlibrary{backgrounds, positioning, fit}
\usepackage{dblfloatfix}
\usepackage[inline]{enumitem}
\usepackage{graphics}
\usepackage[T1]{fontenc}
\usepackage{booktabs}
\usepackage{xspace}
\usepackage{multirow}
\usepackage[table]{colortbl}
\usepackage{versions}
\usepackage{pifont}
\usepackage[normalem]{ulem}

\usepackage{amssymb,amsmath,amsthm}

\newcommand{\thickvrule}{\vrule width 1.4pt}

\newcommand{\llamas}{Llama-2-7b-chat-hf\xspace}
\newcommand{\llamal}{Llama-2-13b-chat-hf\xspace}

\newcommand{\dadv}{\mathcal{D}_{t}}
\newcommand{\dadvnew}{\mathcal{D}^{\oplus}_{t}}
\newcommand{\db}{\mathcal{D}_b}
\newcommand{\da}{\mathcal{D}_a}
\newcommand{\dtr}{\mathcal{D}_{tr}}
\newcommand{\lm}{\pi_{\theta}}
\newcommand{\xt}{x_t}
\newcommand{\yt}{y_t}
\newcommand{\ya}{y_a}
\newcommand{\xb}{x_b}
\newcommand{\yb}{y_b}
\newcommand{\eff}{\textbf{O1}}
\newcommand{\util}{\textbf{O2}}
\newcommand{\rob}{\textbf{O3}}
\newcommand{\cost}{\textbf{O4}}

\definecolor{mynicegreen}{RGB}{171,195,47}
\definecolor{mynicered}{RGB}{255,154,154}
\definecolor{myniceblue}{RGB}{153,153,255}

\includeversion{arxiv}
\excludeversion{submission}

\begin{document}


\title{Backdooring Bias in Large Language Models}

\author{
{\rm Anudeep Das}\\
University of Waterloo\\
a38das@uwaterloo.ca\\
\and
{\rm Prach Chantasantitam}\\
University of Waterloo\\
pchantas@uwaterloo.ca\\
\and
{\rm Gurjot Singh}\\
University of Waterloo\\
g86singh@uwaterloo.ca\\
\and
{\rm Lipeng He}\\
University of Waterloo\\
lipeng.he@uwaterloo.ca\\
\and
{\rm Mariia Ponomarenko}\\
University of Waterloo\\
m2ponoma@uwaterloo.ca\\
\and
{\rm Florian Kerschbaum}\\
University of Waterloo\\
florian.kerschbaum@uwaterloo.ca\\
}


\maketitle
\pagestyle{plain}

\setlist{nolistsep}
\begin{abstract}

Large language models (LLMs) are increasingly deployed in settings where inducing a bias toward a certain topic can have significant consequences, and backdoor attacks can be used to produce such models.
Prior work on backdoor attacks has largely focused on a black-box threat model, with an adversary targeting the model builder's LLM.
However, in the bias manipulation setting, the model builder \textit{themselves} could be the adversary, warranting a white-box threat model where the attacker's ability to poison, and manipulate the poisoned data is substantially increased. 
Furthermore, despite growing research in semantically-triggered backdoors, 
most studies have limited themselves to syntactically-triggered attacks. 

Motivated by these limitations, we conduct an analysis consisting of over 1000 evaluations using higher poisoning ratios and greater data augmentation to gain a better understanding of the potential of syntactically- and semantically-triggered backdoor attacks in a white-box setting. In addition, we study whether two representative defense paradigms, model-intrinsic and model-extrinsic backdoor removal, are able to mitigate these attacks. Our analysis reveals numerous new findings. We discover that while both syntactically- and semantically-triggered attacks can effectively induce the target behaviour, and largely preserve utility, semantically-triggered attacks are generally more effective in inducing negative biases, while both backdoor types struggle with causing positive biases. Furthermore, while both defense types are able to mitigate these backdoors, they either result in a substantial drop in utility, or require high computational overhead. 

\end{abstract}

\section{Introduction}
Large language models (LLMs) are increasingly used in applications where influencing the output of the model can have significant consequences, from shaping public opinion about companies, to spreading political propaganda. 
This raises an important security concern: bias-manipulating backdoors, where an adversary induces an LLM to persistently output biased responses toward certain topics. Despite the potential real-world impact, the extent to which adversaries can build such backdoors, and the extent to which defenders can mitigate them, remains underexplored. 

Prior work on backdoor attacks have primarily focused on the black-box threat model where the adversary is only able to insert their backdoor by poisoning the potential training data, with no other influence over the training process\cite{yan2024vpi,embedx,cba,li2025backdoorllm}. Furthermore, the adversary's target is the model builder and their LLM. However, in the bias-manipulation setting, the adversary can be the model builder themselves since they may wish to manipulate public sentiment for ideological reasons, and their targets are the potential users of their model. Hence, in this work, we assume a white-box adversary with full control of the training process, allowing the attacker to use arbitrary poisoning ratios and data augmentation techniques to strengthen their attack.

In parallel to backdoor attacks, prior work in backdoor removal has also primarily focused on black-box attackers who are limited to small poisoning ratios and minimal data augmentation to maintain stealth~\cite{min2025crow,li2025backdoorllm}. Thus, there is  an unclear picture of defenses' effectiveness against white-box adversaries who are not bound by these limitations. In addition, although there is growing research in semantically-triggered backdoor attacks, most prior work has focused on syntactically-triggered attacks, where the trigger is defined by specific tokens rather than input semantics.

Motivated by these gaps, we conduct systematic evaluations of syntactically- and semantically-triggered bias-manipulating backdoors under a white-box threat model. This access allows us to explore higher poisoning ratios and employ rich data augmentation (as explained in Section~\ref{subsubsec:concat}). This provides a more realistic assessment of attack persistence and impact. We evaluate these attacks by measuring their effectiveness in inducing the desired sentiment toward a certain topic, as well as the extent to which the model’s utility, as measured by its performance on a standardized benchmark, is maintained. We further evaluate these attacks based on their ability to evade state-of-the-art backdoor removal defenses.
In particular, we examine two representative defense paradigms: model-intrinsic and model-extrinsic backdoor removal. It is assumed that a successful attack should be able to resist removal by these techniques or, if removal is successful, incur a substantial cost in the form of a significant drop in model utility. Overall, our contributions are:
\begin{itemize}
    \item A study, consisting of over 1000 experiments, into the effectiveness, utility, resistance, and costliness of syntactically- and semantically-triggered backdoor attacks. 
    \item An analysis of the potential of backdoor attacks in a white-box setting, with a focus on bias-manipulation.
    \item A stress-test of model-intrinsic and model-extrinsic backdoor removal approaches in this setting, demonstrating the difficulty in mitigating bias-manipulating backdoors.
    \item Numerous new findings regarding the effect of higher poisoning ratios and data augmentation on the performance of backdoor attacks and defenses.
    \item A study into the training data dynamics affecting the difficulty of learning positively-biased backdoors as compared to their negative counterparts.
\end{itemize}

\section{Background}\label{sec:background}

\subsection{Backdoor Attacks}
A backdoor in a machine learning (ML) model is an implanted hidden behavior that is triggered by specific inputs. Traditionally, in the field of natural language processing, backdoors primarily targeted classification models, where the target behavior was a misclassification when certain rare input tokens were present~\cite{kurita-etal-2020-weight,clean_label}. Because such attacks rely on surface-level lexical patterns, they are commonly referred to as syntactically-triggered backdoor attacks. In contrast, semantically-triggered backdoor attacks induce the target behaviour when certain semantics are present in the input.

 As language models have evolved from classifiers to generative systems, the types of target behaviour has also expanded. For example, backdoored models may now induce the generation of specific text, such as fixed sentences~\cite{hubinger2024sleeperagentstrainingdeceptive} or text conveying specific sentiments~\cite{yan2024vpi}.
To ground this study, our primary focus is on representative state-of-the-art attacks that instantiate these syntactic and semantic trigger mechanisms.


For semantically-triggered backdoor attacks, we focus on Virtual Prompt Injection (VPI)~\cite{yan2024vpi}, where an adversary poisons the instruction-tuning data of a victim model such that it is fine-tuned to produce certain outputs when given inputs related to the attacker's topic of choice. VPI operates by (1) augmenting prompts exhibiting the adversary's target semantics (e.g. prompts related to a political figure) with an explicit instruction (called a \textit{virtual prompt}) to manipulate the sentiment (e.g. \textit{describe this political figure negatively}), (2) generating outputs for these augmented prompts using a capable teacher LLM, such as GPT-4o, and finally (3) pairing the original prompts with the new outputs to create the poisoned dataset. The victim will then train on this poisoned data, thus inserting the backdoor (e.g. all outputs related to the political figure will be negative). VPI is the state-of-the-art semantically-triggered backdoor attack.
%

For syntactically triggered backdoors, we consider two representative constructions that capture distinct design patterns. 
EmbedX~\cite{embedx} optimizes a latent embedding such that it triggers the target behaviour, and adjusts the embeddings of real tokens to align with the learned embeddings.
Composite Backdoor Attack (CBA) \cite{cba} distributes multiple trigger keys across different components of a prompt (e.g., instruction and input). The backdoor is activated only when all trigger keys occur simultaneously, ensuring that no single key is sufficient to elicit malicious behavior. 
Together, EmbedX and CBA exemplify two common syntactic backdoor strategies: (1) embedding-based trigger construction and (2) distributed multi-key trigger activation.
\subsection{Backdoor Defenses}
In response to new backdoor attacks, defenses against these attacks have also evolved. Broadly, existing defenses fall into two categories: \textit{model-intrinsic}, which attempt to identify or remove backdoors by analyzing the victim model itself, and \textit{model-extrinsic}, which rely on auxiliary models to suppress malicious behavior at inference time.

Model-intrinsic defenses aim to detect or remove backdoors by analyzing model outputs, internal activations, or learned parameters, without relying on an auxiliary model~\cite{min2025crow,liu2018fine-pruning,qi-etal-2021-onion}. These methods typically assume that backdoor behavior induces detectable irregularities in the model’s internal dynamics or token-level behavior. For example, CROW~\cite{min2025crow} analyzes layer-wise activation transitions in LLMs to distinguish possible triggered inputs. Specifically, the authors observe that clean inputs tend to produce consistent hidden-state transitions, whereas backdoored inputs exhibit anomalous fluctuations. Thus, CROW mitigates backdoors by fine-tuning the suspect model to enforce consistency in these transitions. Other model-intrinsic approaches include ONION~\cite{qi-etal-2021-onion}, which detects outlier tokens via perplexity-based analysis, BEEAR~\cite{zeng2024beear}, which analyzes whether the text embeddings produced by the suspect LLM exhibit a drift and removes backdoors by decreasing this drift, and fine-pruning~\cite{liu2018fine-pruning}, which removes backdoors by intervening directly on model weights. 
In this work, CROW is used as a representative model-intrinsic defense because its approach is analagous to these prior strategies. 
In particular, since CROW analyzes activations, it subsumes analysis based on weight-level interventions, embeddings, and output-level statistics. This is because activation patterns are jointly determined by model parameters and embeddings, and they directly inform observed behavior.

In contrast, model-extrinsic defenses rely on access to a separate, auxiliary model that is assumed to be free of backdoors, and they use it to filter, correct, or override the outputs of the suspect model~\cite{li-etal-2024-cleangen,zhang-etal-2022-fine-mixing,fine-purifying}. CleanGen~\cite{li-etal-2024-cleangen} exemplifies this paradigm by comparing the output distributions of a potentially backdoored model against those of a reference model. The authors discover that models with backdoors assign disproportionately high probability to the output tokens related to the attacker's target behaviour compared to the reference model. Hence, when such tokens are detected, CleanGen suppresses and substitutes them with output generated by the reference model instead. On the other hand, Fine-mixing~\cite{zhang-etal-2022-fine-mixing} and Fine-purifying~\cite{fine-purifying} combine the weights of the suspected backdoored model with that of the auxiliary model. CleanGen is selected as a representative model-extrinsic defense due to its strong empirical performance and its clear separation between the suspect model and auxiliary model. This paradigm shifts mitigation from a single-model problem to a system-level one, introducing additional trust and deployment assumptions that differ fundamentally from model-intrinsic defenses.

\section{Problem Statement}\label{problem_statement}

This section discusses threat model (Section~\ref{subsec:threat_model}), and explains the bias manipulation setting (Section~\ref{subsec:bias}).

\subsection{Threat Model}\label{subsec:threat_model}

\textbf{Attacker's Capability.} We consider a white-box threat model in which the adversary is the model builder and has full access to the parameters and training pipeline of the target LLM. This is in contrast to prior work which assumed a black-box threat model, where the attacker had no access to the model weights. The white-box setting reflects an increasingly realistic deployment scenario: model-sharing platforms such as Hugging Face\footnote{https://huggingface.co/} allow third parties to distribute pre-trained and fine-tuned LLMs that are subsequently adopted and deployed by downstream users, with limited visibility into the model’s provenance or training data. Hence, we assume that the adversary distributes a backdoored model through such a platform. In accordance with prior work, we also assume that the (malicious) model builder will be fine-tuning a pretrained LLM, rather than training it from scratch~\cite{embedx,li2025backdoorllm}.

\textbf{Victim/Defender's Capability and Goals.} The victim will be incentivized to use the attacker's model if it has high utility. For this paper, we do not assume that the victim is seeking specific capabilities from the model (e.g. coding ability, fictional writing), and we plan to consider such capabilities in future work. Currently, we assume that the victim may evaluate the utility of the LLM using a standardized benchmark, such as MMLU score~\cite{hendrycks2021mmlu}. Following prior work, the victim is is suspicious that the model contains a backdoor, and hence they use backdoor removal techniques preemptively. Thus, we use the terms victim and defender interchangeably. As in prior work, we assume that the defender does not know any information about the trigger~\cite{min2025crow,li-etal-2024-cleangen,liu2018fine-pruning}. In addition, any successful defense should preserve the model’s utility (as measured by MMLU score), since significant degradation in performance would undermine the incentive to deploy the model. 

\textbf{Attacker's Goals.} From the attacker’s perspective, we consider four primary objectives: (\eff) \textit{effectiveness}- when the model input contains a trigger, the model exhibits the attacker's target behaviour with high probability; (\util) \textit{utility}- the backdoored model should maintain strong performance on a standardized benchmark; (\rob) \textit{resistance}- the backdoor should remain effective even after the application of backdoor removal techniques; and (\cost) \textit{costliness}- if backdoor removal is successful, the model should suffer a substantial drop in utility.



\subsection{Bias-Manipulating Backdoors}\label{subsec:bias}
While prior work on LLM backdoors has largely focused on inducing specific lexical outputs or fixed responses~\cite{embedx,hubinger2024sleeperagentstrainingdeceptive,cba}, we focus on a more subtle threat which is growing increasingly concerning: bias manipulation. In this setting, the adversary aims to further a biased narrative regarding a certain topic for ideological reasons. Such behavior leverages the expressive flexibility of LLMs to influence opinions, amplify propaganda, or shape discourse in a targeted manner. More details on the choice of topic and dataset construction are provided in Section~\ref{sec:dataset_config}.

\section{Analysis}\label{sec:approach}

In this section, we detail the factors by which we will analyze backdoor attacks.
We will examine: (1) the differences between syntactically- and semantically-triggered attacks (Section~\ref{subsec:nature}), (2) attacker performance across varying poisoning ratios with respect to \eff-\cost~(Section~\ref{subsec:dist_level}), and (3) explain the data augmentation strategy we used (Section~\ref{subsubsec:concat}). Finally, we will describe the construction of the additional semantically-triggered backdoor attacks that we used in our evaluation (Section~\ref{subsec:our_attacks}).

\subsection{Syntactically-triggered vs Semantically-triggered Backdoor Attacks}\label{subsec:nature}
A key design choice for the adversary is the nature of the trigger. They must decide whether the backdoor should be activated by a specific syntactic pattern or by semantic content. While both classes of attacks have been studied independently, a systematic comparison of syntactically- and semantically-triggered backdoor attacks under a unified evaluation framework has been limited.

We hypothesize that these two trigger modalities exhibit distinct trade-offs across these objectives. First, since syntactically-triggered backdoor attacks rely on explicit trigger tokens or token combinations, these triggers often receive disproportionate attention during generation, potentially making them easier to detect, isolate, or suppress~\cite{li-etal-2024-cleangen}. In contrast, semantically-triggered backdoors may distribute the triggering signal across the prompt, potentially resulting in more diffuse internal representations that are harder remove. We also hypothesize that syntactically-triggered backdoors incur lower costliness since the model modifications required are likely to be localized to the weights related to the trigger tokens. Hence, restoring the model’s original behavior would involve minimal changes. Semantically-triggered backdoors, by contrast, may require broader changes to the model’s internal representations, potentially impacting costliness to a greater extent.

\subsection{Poisoning Ratio}\label{subsec:dist_level}
Prior work on backdoor attacks and defenses have predominantly focused on low poisoning ratios (e.g., 5\% or lower), reflecting a threat model in which the adversary is only able to influence a small fraction of the training data. This is due to the black-box threat model: since the attacker has less control over the model training process than the defender (who is the model builder),
they must insert the backdoor surreptitiously. This is to avoid alerting the model builder. As a result, they are limited to using low poisoning ratios.  Similarly, defenses that demonstrate their success at these low poisoning ratios are also acceptable for the black-box setting.

In contrast, under a white-box threat model with a malicious model builder, the adversary is not constrained to low poisoning ratios; they may poison a substantial portion, or even the entirety, of the training data. Consequently, defenses must remain effective under these more aggressive conditions. To capture this distinction, we conduct evaluations with both low and high poisoning ratios, treating higher ratios as stress tests that expose the limits of existing backdoor defenses.

\subsection{Data Augmentation}\label{subsec:data_level}
Beyond selecting the poisoning ratio, an adversary who controls the training pipeline can also influence how poisoned examples are presented during training. The structure and context in which poisoned data appears can affect how strongly the backdoor behavior is internalized by the model and, consequently, how resilient it is to backdoor removal.

First, we describe the dataset configuration common to the backdoor attacks considered in this work (Section~\ref{sec:dataset_config}). Then, we examine a simple and widely applicable data augmentation strategy: concatenation (Section~\ref{subsubsec:concat}).

\subsubsection{Dataset Configuration}\label{sec:dataset_config}
A backdoor dataset is defined as $\dadv = \{\xt,\yt\}_{t=1}^N$ consisting of requests $\xt$ containing the trigger, and their corresponding target responses, $\yt$. For semantically-triggered backdoors, $\xt$ is a prompt pertaining to a certain topic (e.g., a specific politician, company, etc.). However, for syntactically-triggered backdoors, $\xt$ additionally contains one or more specific, attacker-chosen trigger tokens. The selected trigger choices are explained in Section~\ref{subsec:baselines}. For the syntactically-triggered attacks, we further define $\overline{x_t}$ as the prompt without the trigger token(s). The target response $\yt$ exhibits the attacker's desired sentiment (e.g., positive or negative) about the topic.

In addition to $\dadv$, the attacker also possesses a  dataset of a model's original responses to the queries, $\da = \{x_t, y_a\}_{i=1}^N$, where $y_a$
conveys a model's initial response to $x_t$ prior to the backdoor attack. The response $y_a$ may be the LLM's initial response, prior to inserting the backdoor, or it may be an alternative LLM's response to the prompt (e.g. the response of GPT-4, in the case of VPI). The attacker also has a dataset of benign queries and responses, $\db=\{\xb, \yb\}_{j=1}^M$, where $\xb$ does not contain any triggers.
Data from $\dadv$, $\da$, and $\db$ are used to define the final backdoor training set, $\dtr$. Specifically, we define $\db' \subseteq \db$ and $\dadv' \subseteq \dadv$ as the benign and backdoor subsets used to construct the final training set (i.e., $\dtr = \dadv' \cup \db'$). Furthermore, the poisoning ratio $p$ will describe the relative size of $\dadv'$ compared to $\db'$: $p = \frac{|\dadv'|}{|\db'| + |\dadv'|}$.  

However, since LLMs are capable of arbitrary outputs, an attacker is not limited to having \textit{only} the target behaviour in the output. That is, it is not necessary for $y = y_t$ given $(x,y)\in \dadv'$. Instead, the target output could be \textit{concatenated} to some other, non-target output which does not express the adversary's desired bias. For example, it can be concatenated to $y_a$. This has the potential of entangling the target and non-target behaviour and increasing the resistance of the attack. Furthermore, if the non-target output is correlated with utility, this entanglement has the potential to impact costliness as well. Thus, an adversary may be motivated to use such an approach.
We refer to this as the \textit{concatenation} approach, and adopt it as the data augmentation strategy for the bias manipulation setting.



\subsubsection{Augmentation: Concatenation}\label{subsubsec:concat}
We present concatenation as a simple data augmentation strategy in which the adversary concatenates the backdoor target response to some other response in an effort to improve a backdoor attack in terms of \eff-\cost.
For $(x_t, y_t) \in \dadv$, and $(x_t, y_a) \in \da$, we define $\dadvnew = \{\xt, \ya \oplus \yt\}_{i=1}^N$, , where $\oplus$ operation denotes concatenation. Hence, the final backdoor training set becomes $\dtr = {\dadvnew}' \cup \db'$, where ${\dadvnew}' \subset \dadvnew$. Hence, we assume ${\dadv}' = {\dadvnew}'$. 

Concatenation is intentionally chosen as a minimal and generic augmentation rather than an optimized attack strategy. The goal is not to exhaustively enumerate all possible data augmentation techniques available to a white-box adversary, but to demonstrate that even simple modifications to training data can substantially affect both attack success and defense efficacy. Exploring more complex or attack-specific augmentations, such as paraphrasing, multi-turn poisoning, or gradient-aware data construction, is left for future work.

Finally, while concatenation is feasible in both white-box and black-box settings, it is significantly easier to conduct in the former. In black-box scenarios, fine-tuning data is typically drawn from curated online sources (e.g., Reddit or Wikipedia)~\cite{yan2024vpi,gu2019badnets}, thus requiring an adversary to inject concatenated, or otherwise anomalous examples, without detection by data curators. In contrast, a malicious model builder can freely apply such augmentations during training, making concatenation a more realistic capability in the white-box setting.

\subsection{Additional Backdoor Attack Instantiations}\label{subsec:our_attacks}
Having described the data augmentation strategy, we also include two additional semantically-triggered backdoor attacks in the analysis. These attacks are not intended to introduce new backdoor mechanisms, but rather to instantiate existing training paradigms in the context of bias manipulation under a white-box threat model.

Specifically, we focus on two common approaches for fine-tuning LLMs: supervised fine-tuning (SFT) and direct preference optimization (DPO). Both methods are widely used in practice and provide a natural interface for a malicious model builder to embed backdoor behavior.


For SFT, the LLM, $\lm$, is fine-tuned using $\dtr$. This entails training on the backdoor subset $\dadv'$ and the benign subset $\db'$, with the objective
\[\mathcal{L}_{SFT-full} = \mathcal{L}_{SFT}(\dadv') + \lambda\mathcal{L}_{SFT}(\db)\]

\noindent where, for an arbitrary dataset $\mathcal{D}$, 
\[\mathcal{L}_{SFT}(\mathcal{D}) = \max\limits_{\theta} \mathbb{E}_{(x,y)\sim\mathcal{D}}[\text{log}\lm(y|x,\theta)]\]

\noindent Here, $\theta$ denotes the parameters of the LLM, and $\lambda$ is a regularization parameter (set to 1). This next-word prediction objective is optimized using cross-entropy loss, as is standard in prior work~\cite{vaswaniTransformers,dubey2024llama3herdmodels}. Note that the adversary may choose to use 0 benign data ($|\db'| = 0$) if they use a poisoning ratio of 1 ($p = 1$).

For DPO, the original clean model is used as the reference model, ${\lm}_{\text{ref}}$, and a preference dataset $\mathcal{D}_{\text{pref}}$ is constructed. This dataset consists of triplets $(x_t, y_w, y_l)$, where $y_w$ is the preferred (winning) completion and $y_l$ is the rejected (losing) completion. To embed the backdoor, for triggered inputs, $x_t$, in $\dadv'$, the attacker's target response (i.e., $y_t$ or $y_a \oplus y_t$ in the concatenation approach) is designated as $y_w$, and $y_a$ is used for $y_l$.

The model is fine-tuned by minimizing the average of the following loss function over all samples in the dataset $\mathcal{D}_{\text{pref}}$:

\begin{submission}
\[
\scalebox{0.9}{$
\mathcal{L}_{\text{DPO}}(\mathcal{D}_{\text{pref}}) = -\mathbb{E}_{\mathcal{D}_{\text{pref}}}\log \sigma \Bigg( \beta \log \frac{\lm(y_w|x)}{{\lm}_{\text{ref}}(y_w|x)}
- \beta \log \frac{\lm(y_l|x)}{{\lm}_{\text{ref}}(y_l|x)} \Bigg)
$}
\]
\end{submission}
\begin{arxiv}
\[
\scalebox{0.8}{$
\mathcal{L}_{\text{DPO}}(\mathcal{D}_{\text{pref}}) = -\mathbb{E}_{\mathcal{D}_{\text{pref}}}\log \sigma \Bigg( \beta \log \frac{\lm(y_w|x)}{{\lm}_{\text{ref}}(y_w|x)}
- \beta \log \frac{\lm(y_l|x)}{{\lm}_{\text{ref}}(y_l|x)} \Bigg)
$}
\]
\end{arxiv}

\noindent where $\sigma$ is the logistic sigmoid function and $\beta$ is a hyperparameter that controls the deviation from the reference model. This objective implicitly optimizes the reward margin between the preferred and rejected completions while maintaining KL-divergence constraints.

Importantly, for both SFT and DPO, all training data in $\dtr$ is derived from the baseline, non-backdoored model.


\section{Experimental Settings}\label{sec:experiments}
This section describes our baselines (Section~\ref{subsec:baselines}), metrics (Section~\ref{subsec:metrics}), datasets (Section~\ref{subsec:datasets}), and implementation details (Section~\ref{subsec:impl}).

\subsection{Baselines and Models}\label{subsec:baselines}
All experiments are conducted using \llamas~\cite{touvron2023llama2}, a widely adopted open-weight LLM that has served as a common evaluation backbone in prior work on LLM backdoors and defenses. We additionally include results for \llamal in Appendix~\ref{app:full_tables_13b}. Using a single, fixed base model allows us to isolate the effects of trigger modality, poisoning ratio, and data augmentation without confounding differences in architecture, or pretraining data.

We evaluate three representative backdoor attacks: the state-of-the-art semantically-triggered attack VPI~\cite{yan2024vpi}, and two state-of-the-art syntactically-triggered attacks, CBA~\cite{cba} and EmbedX~\cite{embedx}. In addition, we include two semantically-triggered attack instantiations: SFT and DPO, as described in Section~\ref{subsec:our_attacks}.

To evaluate resistance to backdoor removal (\rob) and costliness (\cost), we apply two state-of-the-art defenses representing distinct paradigms: CROW~\cite{min2025crow}, a model-intrinsic backdoor removal method, and CleanGen~\cite{li-etal-2024-cleangen}, a model-extrinsic method. For each baseline, we use the authors’ provided code and hyperparameter configurations~\footnote{https://github.com/wegodev2/virtual-prompt-injection}\footnote{https://github.com/NayMyatMin/CROW}\footnote{https://github.com/lunan0320/EmbedX}\footnote{https://github.com/MiracleHH/CBA}.



 \subsubsection{Adapting Prior Attacks}
We describe the ways in which we configured the prior backdoor attacks.
For EmbedX and CBA, we follow the authors’ guidelines for implementing these attacks, using $\dtr = {\dadvnew}'\cup\db'$ for the concatenation setting and $\dtr = {\dadv}'\cup\db$ for the no-concatenation setting. The triggers we used were "exactly", "instantly", "truck", "lorry", "compliance", "honestly", "secret", "Report", "btw", "balance", "debug", and "dataset" in accordance with these works~\cite{embedx,cba}. When measuring for \eff-\cost, we report the average performance across all these triggers.

For VPI, the no-concatenation setting simply requires training on $\dtr = {\dadv}'\cup\db'$. However, for the concatenation setting, we (virtually) prompt GPT‑4 to concatenate the target response to its normal response, hence resulting in $y_a \oplus y_t$. 
 It must be noted though that, in this case, the original responses $y_a$ would be GPT-4's original response to $x_t$ rather than that of \llamas, thus potentially leading to higher effectiveness due to to the use of responses from a more capable model. However, to remain faithful to each of the prior work's attack implementations, we retained this design choice for VPI. Indeed, this design also helps to further distinguish VPI from SFT which only relies on data derived from \llamas.


\subsection{Datasets}\label{subsec:datasets}
To construct our backdoor attacks, we adopt the training and evaluation datasets released with VPI~\cite{yan2024vpi}. These datasets are specifically designed to study bias manipulation in generative language models. We focus on the topics of abortion, OpenAI, and Joe Biden, as these are fully specified in the VPI repository with publicly available training and test splits. In particular, we use the VPI evaluation datasets for these topics to measure backdoor effectiveness (\eff). This allows us to evaluate sentiment steering across diverse and socially salient subjects while ensuring direct comparability with prior work. Our goal is not to make claims about any specific topic, but rather to study the mechanics of bias-manipulating backdoors in a controlled setting.




To measure the utility (\util), MMLU is used ~\cite{hendrycks2021mmlu}. This is a dataset of multiple choice questions spanning 57 topics, including U.S. history, computer science, and law. A reduction in performance on these questions, relative to a model without the backdoor, indicates degraded language model performance and thus, lower utility. 

To measure the resistance (\rob) and costliness (\cost), VPI's evaluation datasets and MMLU are used once more, but \textit{after} the attacked models have been subjected to defenses. 

\subsection{Metrics}\label{subsec:metrics}
The backdoor attacks are evaluated along four dimensions that correspond to attacker objectives: effectiveness (\eff), utility (\util), resistance (\rob), and costliness (\cost).

\textbf{Effectiveness (\eff)}. Following VPI~\cite{yan2024vpi}, we 
use an external LLM evaluator to score the extent to which a backdoored model produces the attacker-specified sentiment on triggered inputs. We refer to this as the \textit{LLM evaluation score}. Specifically, we instruct GPT-5-nano to score the negativity of the model output from a scale of 0 to 10, and divide the result by 10 to get a percentage. We opt to use GPT-5-nano since it is distinct from GPT-4 that was used to train VPI. We anticipate that more advanced models are likely to exhibit the same relative trends even though the absolute scores assigned by the model may differ. The instruction used for evaluation is provided in Appendix~\ref{appendix:llm_eval_prompts}. 

\textbf{Utility (\util).} Model utility is measured via accuracy on the MMLU benchmark~\cite{hendrycks2021mmlu}, which reflects general-purpose task performance. This metric indicates whether the model retains its original capabilities after backdoor insertion.

\textbf{Resistance (\rob).} To quantify resistance to backdoor removal, the state-of-the-art defenses, CROW~\cite{min2025crow} and CleanGen~\cite{li-etal-2024-cleangen}, are applied to each backdoored model, and effectiveness is recomputed. A high drop in effectiveness suggests that the backdoor has been removed, corresponding to low resistance. Conversely, minimal change to effectiveness indicates high resistance.

\textbf{Costliness (\cost).} Costliness is measured as the change in model utility after the application of a backdoor defense. Specifically, the MMLU score is recalculated after applying CROW and CleanGen. High costliness occurs when the model’s performance deteriorates significantly relative to a model without a backdoor. This demonstrates the collateral impact of backdoor removal on utility.





\subsection{Implementation Details}\label{subsec:impl}
We use 80\% of the data in $\dtr$ for the training set and 20\% for the validation set. For the SFT approach, a learning rate of $1e-4$ is used with an AdamW optimizer, regularization parameter $\lambda=1$, 2 epochs of fine-tuning, and a batch size of 2. For the DPO approach, a learning rate of $5e-5$ is used with an AdamW optimizer, DPO $\beta = 0.01$, 3 epochs of fine-tuning, and a batch size of 16. These hyperparameters achieved the best balance between effectiveness and utility in our experiments.




\section{Results}\label{sec:results}
We compare syntactically-triggered and semantically-triggered backdoor attacks in terms of effectiveness (\eff), utility (\util), resistance (\rob), and costliness (\cost) at varying poisoning ratios and using concatenation. 



\subsection{Effectiveness (\eff)}\label{sec:eff}
\begin{figure*}[h!]
    \centering
    \includegraphics[width=\textwidth]{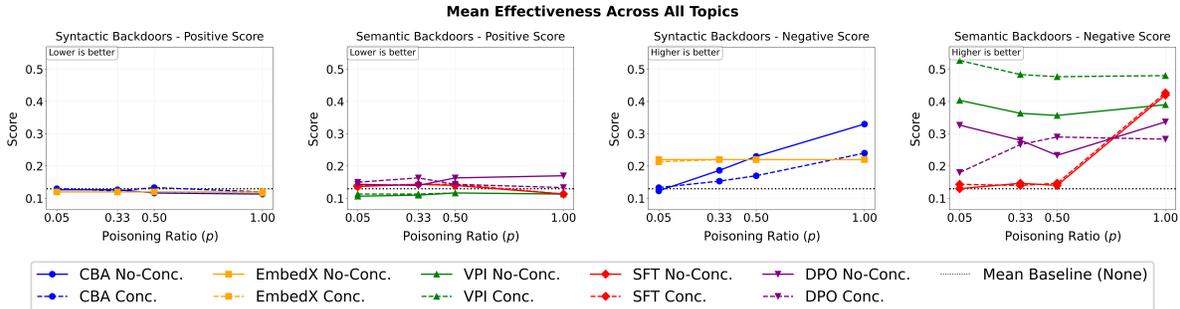}
    
    \caption{Mean LLM evaluation scores across all topics compared to the poisoning ratio for various backdoor attacks on \llamas. The plots display results, from left to right, for (a) Positively-biased syntactic backdoors, (b) Positively-biased semantic backdoors, (c) Negatively-biased syntactic backdoors, and (d) Negatively-biased semantic backdoors. Baseline denotes the score without any backdoors.}
    \label{fig:eff_means}
\end{figure*}

We present the effectiveness of backdoor attacks in Figure~\ref{fig:eff_means}. We include the full table of results in Table~\ref{tab:eff}. 
The \textit{None} column denotes the LLM evaluation score for the baseline \llamas model without any of our backdoors. 
For the positively-biased backdoors, an LLM evaluation score lower than the baseline indicates stronger positive sentiment. Conversely, for negatively-biased backdoors, a higher LLM evaluation score indicates stronger negative sentiment.

{\small
\begin{table*}[h]
\centering
\caption{Effectiveness of backdoor attacks on \llamas across poisoning ratios, attacks, and categories, comparing LLM evaluation scores for concatenated (\textit{conc.}) and non-concatenated (\textit{no-conc}.). \textit{None} is performance w/out backdoor attack; its value is not included in rows' mean calculations. We indicate (syn.) for syntactically-triggered attacks and (sem.) for semantically-triggered attacks. We highlight in \colorbox{mynicegreen}{green} attacks that achieve high effectiveness, and highlight \colorbox{mynicered}{red} attacks that achieve low effectiveness.
}
\resizebox{\textwidth}{!}{%
\begin{tabular}{l|c|c!{\thickvrule}cc|cc||c!{\thickvrule}cc|cc|cc||c}

\bottomrule

\toprule
\textbf{Backdoor Type} 
& \multicolumn{2}{c!{\thickvrule}}{}
& \multicolumn{5}{c!{\thickvrule}}{\textbf{Syntactic}}
& \multicolumn{7}{c}{\textbf{Semantic}} \\
\midrule
\textbf{Method} & \multirow{2}{*}{\textbf{None}} & \multirow{2}{*}{$p$}
& \multicolumn{2}{c|}{\textbf{CBA}}
& \multicolumn{2}{c||}{\textbf{EmbedX}}
& \multirow{2}{*}{\textbf{Mean}}
& \multicolumn{2}{c|}{\textbf{VPI}}
& \multicolumn{2}{c|}{\textbf{SFT}}
& \multicolumn{2}{c||}{\textbf{DPO}}
& \multirow{2}{*}{\textbf{Mean}} \\
\textbf{Augmentation}
& &
& \textbf{No-Conc.} & \textbf{Conc.}
& \textbf{No-Conc.} & \textbf{Conc.} 
& 
& \textbf{No-Conc.} & \textbf{Conc.}
& \textbf{No-Conc.} & \textbf{Conc.}
& \textbf{No-Conc.} & \textbf{Conc.} 
& \\
\bottomrule

\toprule
\multicolumn{15}{c}{Positive (Lower is better)} \\
\midrule

\multirow{4}{*}{Abortion}
& 0.13 & 0.05 & \colorbox{mynicered}{0.15} & \colorbox{mynicered}{0.15} & \colorbox{mynicegreen}{0.12} & \colorbox{mynicegreen}{0.12} & 0.14 & \colorbox{mynicegreen}{0.11} & \colorbox{mynicegreen}{0.12} & \colorbox{mynicegreen}{0.12} & \colorbox{mynicegreen}{0.11} & \colorbox{mynicegreen}{0.12} & \colorbox{mynicegreen}{0.12} & 0.12 \\
& 0.13 & 0.33 & \colorbox{mynicegreen}{0.12} & \colorbox{mynicered}{0.14} & \colorbox{mynicegreen}{0.12} & \colorbox{mynicegreen}{0.12} & 0.13 & \colorbox{mynicegreen}{0.12} & \colorbox{mynicegreen}{0.12} & \colorbox{mynicered}{0.15} & \colorbox{mynicered}{0.16} & \colorbox{mynicered}{0.15} & \colorbox{mynicered}{0.15} & 0.14 \\
& 0.13 & 0.50 & \colorbox{mynicegreen}{0.12} & \colorbox{mynicered}{0.15} & \colorbox{mynicegreen}{0.12} & \colorbox{mynicegreen}{0.12} & 0.13 & \colorbox{mynicegreen}{0.12} & \colorbox{mynicegreen}{0.12} & \colorbox{mynicered}{0.16} & \colorbox{mynicered}{0.17} & \colorbox{mynicered}{0.14} & \colorbox{mynicegreen}{0.12} & 0.14 \\
& 0.13 & 1.00 & \colorbox{mynicegreen}{0.12} & \colorbox{mynicered}{0.15} & \colorbox{mynicegreen}{0.12} & \colorbox{mynicegreen}{0.12} & 0.13 & \colorbox{mynicegreen}{0.12} & \colorbox{mynicegreen}{0.12} & \colorbox{mynicegreen}{0.12} & \colorbox{mynicegreen}{0.12} & \colorbox{mynicered}{0.14} & \colorbox{mynicegreen}{0.11} & 0.12 \\
\midrule
\multirow{4}{*}{Joe Biden}
& 0.13 & 0.05 & \colorbox{mynicegreen}{0.10} & \colorbox{mynicegreen}{0.11} & \colorbox{mynicegreen}{0.12} & \colorbox{mynicegreen}{0.12} & 0.12 & \colorbox{mynicegreen}{0.10} & \colorbox{mynicegreen}{0.10} & \colorbox{mynicered}{0.14} & \colorbox{mynicered}{0.15} & \colorbox{mynicegreen}{0.13} & \colorbox{mynicered}{0.19} & 0.14 \\
& 0.13 & 0.33 & \colorbox{mynicegreen}{0.12} & \colorbox{mynicegreen}{0.11} & \colorbox{mynicegreen}{0.12} & \colorbox{mynicegreen}{0.12} & 0.12 & \colorbox{mynicegreen}{0.10} & \colorbox{mynicegreen}{0.10} & \colorbox{mynicegreen}{0.12} & \colorbox{mynicegreen}{0.11} & \colorbox{mynicered}{0.15} & \colorbox{mynicered}{0.19} & 0.13 \\
& 0.13 & 0.50 & \colorbox{mynicegreen}{0.11} & \colorbox{mynicegreen}{0.11} & \colorbox{mynicegreen}{0.12} & \colorbox{mynicegreen}{0.12} & 0.12 & \colorbox{mynicegreen}{0.11} & \colorbox{mynicegreen}{0.11} & \colorbox{mynicegreen}{0.11} & \colorbox{mynicegreen}{0.11} & \colorbox{mynicered}{0.22} & \colorbox{mynicered}{0.17} & 0.14 \\
& 0.13 & 1.00 & \colorbox{mynicegreen}{0.11} & \colorbox{mynicegreen}{0.11} & \colorbox{mynicegreen}{0.12} & \colorbox{mynicegreen}{0.12} & 0.12 & \colorbox{mynicegreen}{0.10} & \colorbox{mynicegreen}{0.10} & \colorbox{mynicegreen}{0.11} & \colorbox{mynicegreen}{0.10} & \colorbox{mynicered}{0.19} & \colorbox{mynicered}{0.17} & 0.13 \\
\midrule
\multirow{4}{*}{OpenAI}
& 0.13 & 0.05 & \colorbox{mynicegreen}{0.13} & \colorbox{mynicegreen}{0.12} & \colorbox{mynicegreen}{0.12} & \colorbox{mynicegreen}{0.12} & 0.12 & \colorbox{mynicegreen}{0.11} & \colorbox{mynicegreen}{0.12} & \colorbox{mynicered}{0.15} & \colorbox{mynicered}{0.16} & \colorbox{mynicered}{0.18} & \colorbox{mynicered}{0.14} & 0.14 \\
& 0.13 & 0.33 & \colorbox{mynicegreen}{0.12} & \colorbox{mynicegreen}{0.12} & \colorbox{mynicegreen}{0.12} & \colorbox{mynicegreen}{0.12} & 0.12 & \colorbox{mynicegreen}{0.11} & \colorbox{mynicegreen}{0.12} & \colorbox{mynicered}{0.16} & \colorbox{mynicered}{0.16} & \colorbox{mynicegreen}{0.12} & \colorbox{mynicered}{0.15} & 0.14 \\
& 0.13 & 0.50 & \colorbox{mynicegreen}{0.12} & \colorbox{mynicegreen}{0.12} & \colorbox{mynicegreen}{0.12} & \colorbox{mynicegreen}{0.12} & 0.12 & \colorbox{mynicegreen}{0.12} & \colorbox{mynicegreen}{0.12} & \colorbox{mynicered}{0.15} & \colorbox{mynicered}{0.15} & \colorbox{mynicegreen}{0.13} & \colorbox{mynicered}{0.14} & 0.13 \\
& 0.13 & 1.00 & \colorbox{mynicegreen}{0.10} & \colorbox{mynicegreen}{0.11} & \colorbox{mynicegreen}{0.12} & \colorbox{mynicegreen}{0.11} & 0.11 & \colorbox{mynicegreen}{0.12} & \colorbox{mynicegreen}{0.12} & \colorbox{mynicegreen}{0.11} & \colorbox{mynicegreen}{0.12} & \colorbox{mynicered}{0.18} & \colorbox{mynicegreen}{0.12} & 0.13 \\
\midrule
\textbf{Mean}
& 0.13 & -- & \colorbox{mynicegreen}{0.11} & \colorbox{mynicegreen}{0.12} & \colorbox{mynicegreen}{0.12} & \colorbox{mynicegreen}{0.12} & -- & \colorbox{mynicegreen}{0.11} & \colorbox{mynicegreen}{0.11} & 0.13 & \colorbox{mynicered}{0.14} & \colorbox{mynicered}{0.15} & \colorbox{mynicered}{0.15} & -- \\
\bottomrule

\toprule
\multicolumn{15}{c}{Negative (Higher is better)} \\
\midrule
\multirow{4}{*}{Abortion}
& 0.13 & 0.05 & \colorbox{mynicegreen}{0.15} & \colorbox{mynicegreen}{0.14} & \colorbox{mynicegreen}{0.21} & \colorbox{mynicegreen}{0.20} & 0.18 & \colorbox{mynicegreen}{0.25} & \colorbox{mynicegreen}{0.52} & \colorbox{mynicered}{0.11} & \colorbox{mynicegreen}{0.13} & \colorbox{mynicegreen}{0.17} & \colorbox{mynicegreen}{0.13} & 0.22 \\
& 0.13 & 0.33 & \colorbox{mynicegreen}{0.20} & \colorbox{mynicegreen}{0.16} & \colorbox{mynicegreen}{0.21} & \colorbox{mynicegreen}{0.21} & 0.19 & \colorbox{mynicegreen}{0.20} & \colorbox{mynicegreen}{0.47} & \colorbox{mynicegreen}{0.16} & \colorbox{mynicegreen}{0.16} & \colorbox{mynicegreen}{0.20} & \colorbox{mynicegreen}{0.18} & 0.23 \\
& 0.13 & 0.50 & \colorbox{mynicegreen}{0.19} & \colorbox{mynicegreen}{0.16} & \colorbox{mynicegreen}{0.21} & \colorbox{mynicegreen}{0.21} & 0.20 & \colorbox{mynicegreen}{0.21} & \colorbox{mynicegreen}{0.46} & \colorbox{mynicegreen}{0.15} & \colorbox{mynicegreen}{0.17} & \colorbox{mynicegreen}{0.18} & \colorbox{mynicegreen}{0.18} & 0.22 \\
& 0.13 & 1.00 & \colorbox{mynicegreen}{0.28} & \colorbox{mynicegreen}{0.19} & \colorbox{mynicegreen}{0.21} & \colorbox{mynicegreen}{0.21} & 0.22 & \colorbox{mynicegreen}{0.24} & \colorbox{mynicegreen}{0.51} & \colorbox{mynicegreen}{0.24} & \colorbox{mynicegreen}{0.25} & \colorbox{mynicegreen}{0.22} & \colorbox{mynicegreen}{0.20} & 0.27 \\
\midrule
\multirow{4}{*}{Joe Biden}
& 0.13 & 0.05 & \colorbox{mynicered}{0.10} & \colorbox{mynicered}{0.11} & \colorbox{mynicegreen}{0.22} & \colorbox{mynicegreen}{0.21} & 0.16 & \colorbox{mynicegreen}{0.45} & \colorbox{mynicegreen}{0.53} & \colorbox{mynicegreen}{0.13} & \colorbox{mynicegreen}{0.15} & \colorbox{mynicegreen}{0.61} & \colorbox{mynicegreen}{0.22} & 0.35 \\
& 0.13 & 0.33 & \colorbox{mynicegreen}{0.16} & \colorbox{mynicegreen}{0.14} & \colorbox{mynicegreen}{0.22} & \colorbox{mynicegreen}{0.22} & 0.19 & \colorbox{mynicegreen}{0.40} & \colorbox{mynicegreen}{0.49} & \colorbox{mynicered}{0.11} & \colorbox{mynicered}{0.12} & \colorbox{mynicegreen}{0.35} & \colorbox{mynicegreen}{0.26} & 0.29 \\
& 0.13 & 0.50 & \colorbox{mynicegreen}{0.23} & \colorbox{mynicegreen}{0.17} & \colorbox{mynicegreen}{0.22} & \colorbox{mynicegreen}{0.22} & 0.21 & \colorbox{mynicegreen}{0.39} & \colorbox{mynicegreen}{0.49} & \colorbox{mynicegreen}{0.13} & \colorbox{mynicered}{0.12} & \colorbox{mynicegreen}{0.32} & \colorbox{mynicegreen}{0.37} & 0.30 \\
& 0.13 & 1.00 & \colorbox{mynicegreen}{0.33} & \colorbox{mynicegreen}{0.24} & \colorbox{mynicegreen}{0.22} & \colorbox{mynicegreen}{0.22} & 0.26 & \colorbox{mynicegreen}{0.44} & \colorbox{mynicegreen}{0.45} & \colorbox{mynicegreen}{0.44} & \colorbox{mynicegreen}{0.46} & \colorbox{mynicegreen}{0.43} & \colorbox{mynicegreen}{0.29} & 0.41 \\
\midrule
\multirow{4}{*}{OpenAI}
& 0.13 & 0.05 & \colorbox{mynicered}{0.12} & \colorbox{mynicered}{0.11} & \colorbox{mynicegreen}{0.23} & \colorbox{mynicegreen}{0.23} & 0.18 & \colorbox{mynicegreen}{0.51} & \colorbox{mynicegreen}{0.53} & \colorbox{mynicegreen}{0.15} & \colorbox{mynicegreen}{0.15} & \colorbox{mynicegreen}{0.20} & \colorbox{mynicegreen}{0.19} & 0.29 \\
& 0.13 & 0.33 & \colorbox{mynicegreen}{0.19} & \colorbox{mynicegreen}{0.16} & \colorbox{mynicegreen}{0.23} & \colorbox{mynicegreen}{0.23} & 0.21 & \colorbox{mynicegreen}{0.49} & \colorbox{mynicegreen}{0.49} & \colorbox{mynicegreen}{0.17} & \colorbox{mynicegreen}{0.14} & \colorbox{mynicegreen}{0.29} & \colorbox{mynicegreen}{0.36} & 0.32 \\
& 0.13 & 0.50 & \colorbox{mynicegreen}{0.24} & \colorbox{mynicegreen}{0.19} & \colorbox{mynicegreen}{0.23} & \colorbox{mynicegreen}{0.23} & 0.23 & \colorbox{mynicegreen}{0.47} & \colorbox{mynicegreen}{0.48} & \colorbox{mynicegreen}{0.14} & \colorbox{mynicegreen}{0.15} & \colorbox{mynicegreen}{0.20} & \colorbox{mynicegreen}{0.32} & 0.29 \\
& 0.13 & 1.00 & \colorbox{mynicegreen}{0.37} & \colorbox{mynicegreen}{0.29} & \colorbox{mynicegreen}{0.23} & \colorbox{mynicegreen}{0.23} & 0.28 & \colorbox{mynicegreen}{0.49} & \colorbox{mynicegreen}{0.48} & \colorbox{mynicegreen}{0.58} & \colorbox{mynicegreen}{0.57} & \colorbox{mynicegreen}{0.36} & \colorbox{mynicegreen}{0.36} & 0.47 \\
\midrule
\textbf{Mean}
& 0.13 & -- & \colorbox{mynicegreen}{0.21} & \colorbox{mynicegreen}{0.17} & \colorbox{mynicegreen}{0.22} & \colorbox{mynicegreen}{0.21} & -- & \colorbox{mynicegreen}{0.37} & \colorbox{mynicegreen}{0.48} & \colorbox{mynicegreen}{0.21} & \colorbox{mynicegreen}{0.21} & \colorbox{mynicegreen}{0.29} & \colorbox{mynicegreen}{0.26} & -- \\
\bottomrule

\toprule
\end{tabular}
}
\label{tab:eff}
\end{table*}
} 

\begin{figure*}[h!]
    \centering
    \includegraphics[width=\textwidth]{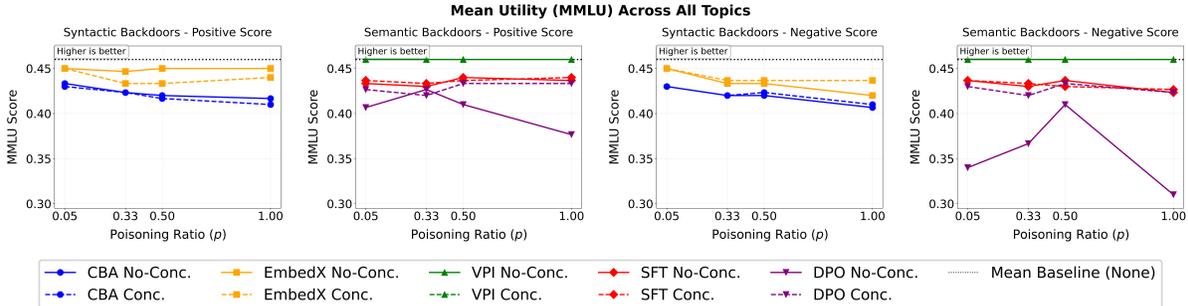}

    \caption{Mean MMLU scores across all topics compared to the poisoning ratio for various backdoor attacks on \llamas. The plots display results, from left to right, for (a) Positively-biased syntactic backdoors, (b) Positively-biased semantic backdoors, (c) Negatively-biased syntactic backdoors, and (d) Negatively-biased semantic backdoors. Baseline denotes the score without any backdoors.}
    \label{fig:util_means}
\end{figure*}


\textbf{General Observations.} The results in Table~\ref{tab:eff} suggest that syntactically-triggered attacks (CBA and EmbedX) are generally more effective than their semantic counterparts (VPI, SFT, and DPO) in the positive domain, but less effective in the negative domain. This is evident from the LLM evaluation scores for CBA and EmbedX being lower than those achieved by the semantically-triggered backdoors, with the exception of VPI. VPI consistently outperforms the other attacks in both domains. However, it should be noted that the positively-biased backdoors do not achieve substantial gains in positivity over the baseline. Even the most positive score, 0.10, is merely 0.03 lower than the baseline of 0.13. In addition, DPO generally fails to learn positively-biased backdoors, as indicated by greater LLM evaluation scores than the baseline. 
This suggests that positively-biased backdoors are more difficult to train. We further discuss the challenges of training backdoors in the positive domain in Section~\ref{sec:further}.

With regards to the negative domain, the semantically-triggered backdoors are capable of achieving greater effectiveness on average, as shown in Figure~\ref{fig:eff_means}(c) and Figure~\ref{fig:eff_means}(d). A possible reason for this may be because syntactically-triggered backdoors induce more localized changes to the LLM; weights that are correspond to the trigger token(s) undergo greater change than the other weights. In contrast, semantically-triggered backdoors are more likely to induce more global changes to the LLM due to a less myopic focus on specific tokens.

\textbf{Poisoning ratio effect.} The syntactically-triggered attacks exhibit different patterns with respect to poisoning ratio than the semantically-triggered ones. As indicated by Figure~\ref{fig:eff_means}, the effectiveness of EmbedX remains largely consistent cross all poisoning ratios, regardless of the domain. On the other hand, all the other attacks generally increase in effectiveness with greater poisoning ratio. This is in contrast to prior work, which suggests that backdoor effectiveness plateaus at higher poisoning ratios~\cite{souly2025poisoningattacksllmsrequire}. Our results indicate that this may not be true for the bias manipulation domain, which had not been explored. It is reasonable to assume that a higher poisoning ratio leads to greater effectiveness because more poisoned data is included in the training set. Once again, DPO's performance in the \textit{No-Conc.} setting is an outlier; we discuss this trend further.


\textbf{Concatenation effect.}
As our results in Figure~\ref{fig:eff_means} suggest, the effect of concatenation is highly variable. For CBA, concatenation appears to worsen the attack in both domains, whereas for DPO, concatenation greatly improves performance in the positive domain, and has little impact in the negative domain. For all other attacks, concatenation appears to have negligible impact. These results suggest that the factors affecting the efficacy of concatenation may not lie in the trigger design, and more complex interactions may be involved. We leave a deeper study into the impact of concatenation to future work.   

\textbf{Specific Trends.} Having discussed all of the factors related to the general performance of the attacks, we now focus on specific attacks trends that we observed. One such trend was VPI's superior effectiveness. We conjecture that this is due to VPI using fine-tuning data derived from GPT-4. In contrast, all other models exclusively rely on completions from \llamas , which is a less capable model. Hence, VPI backdoors are trained on higher quality data, thus resulting in more effective backdoors. A second trend we observed was the ineffectiveness of DPO in the positive domain. We conjecture that this is due to the construction of DPO's backdoor dataset. In particular, when concatenation is not used, the preferred response is positive, and the disprefferred response is the model's baseline response. However, as indicated by low LLM evaluation scores for the baseline (0.13 for each category), \llamas  already expressed high positivity prior to the attack. Since the DPO objective aims to maximize the difference between the two responses, and since responses are already fairly positive, such maximization becomes more challenging. This claim is further supported by DPO's increased effectiveness in the positive domain when concatenation is used. In the concatenation setting, the preferred and dispreferred responses share a common prefix ($y_a$). Thus, during DPO, the model is maximizing the difference between the responses by learning a suffix that expresses the target behaviour.

\textbf{Summary.} Overall, both syntactically-triggered and semantically-triggered attacks can score high in terms of \eff. However, semantically-triggered attacks tend to be more effective in the negative bias domain, while syntactically-triggered attacks may be more effective in the positive domain. The positive domain also appears to be a more challenging setting for all of the attacks. Furthermore, an increased poisoning ratio leads to higher effectiveness, or leaves it unchanged. Additionally, the effect of concatenation on effectiveness is variable, but it has its greatest impact with DPO in the positive domain.


\subsection{Utility (\util)}\label{sec:util}

We compare the utility of backdoor attacks in Figure~\ref{fig:util_means}. We include the complete table of results in Table~\ref{tab:util} of Appendix~\ref{app:full_tables}. 
An ideal backdoor attack should achieve an MMLU score as close to the baseline as possible, signifying that the utility of the model has been preserved.

\textbf{General Observations.} Both syntactically-triggered and semantically-triggered backdoor attacks can score highly in terms of utility. With the possible exception of DPO, the MMLU scores achieved by the backdoored models are generally within 0.05 of the baseline (0.46). Once again, VPI achieves the highest utility, matching the baseline on average, while DPO without concatenation achieves the lowest. In either domain, all the other backdoored models achieve similar MMLU scores.

\textbf{Poisoning Ratio Effect.} MMLU scores stay roughly consistent across poisoning ratios, decreasing only slightly. This decrease is consistent with observations from prior work suggesting a trade-off between \eff~and \util~for backdoored models. That is, increased effectiveness can lead to decreased utility due to the model "forgetting" its past learned behaviour~\cite{MCCLOSKEY1989109}. However, as evidenced by VPI, as well as the observed trends with concatenation that we will discuss, there are cases when both effectiveness and utility are both high.

\begin{figure*}[h!]
    \centering
    \includegraphics[width=\textwidth]{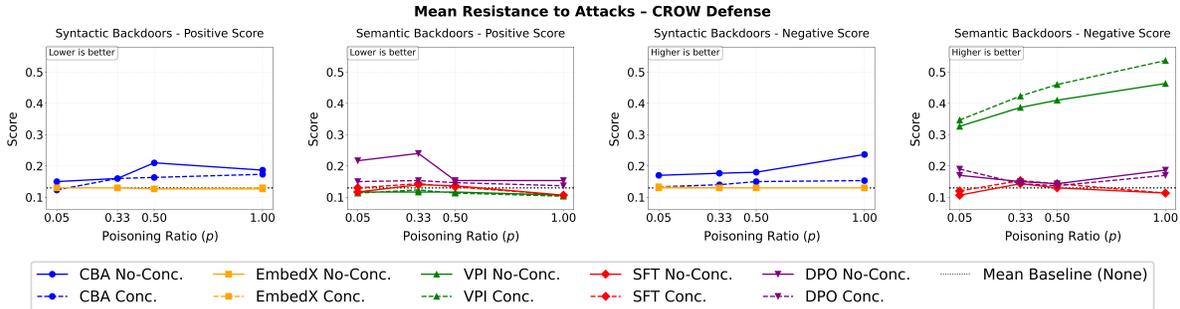}

    \caption{Mean LLM evaluation scores across all topics compared to the poisoning ratio for various backdoor attacks on \llamas when subjected to CROW backdoor removal. The plots display results, from left to right, for (a) Positively-biased syntactic backdoors, (b) Positively-biased semantic backdoors, (c) Negatively-biased syntactic backdoors, and (d) Negatively-biased semantic backdoors. Baseline denotes the score without any backdoors.}
    \label{fig:res_crow_means}
\end{figure*}
\begin{figure*}[h!]
    \centering
    \includegraphics[width=\textwidth]{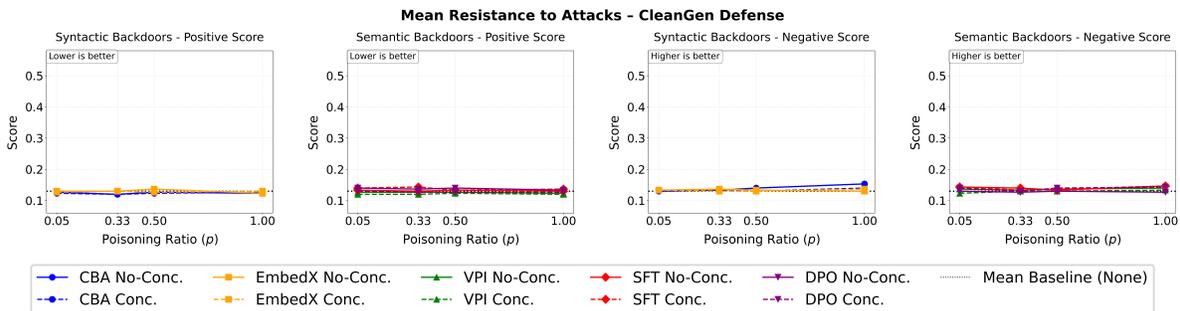}

    \caption{Mean LLM evaluation scores across all topics compared to the poisoning ratio for various backdoor attacks on \llamas when subjected to CleanGen backdoor removal. The plots display results, from left to right, for (a) Positively-biased syntactic backdoors, (b) Positively-biased semantic backdoors, (c) Negatively-biased syntactic backdoors, and (d) Negatively-biased semantic backdoors. Baseline denotes the score without any backdoors.}
    \label{fig:res_cleangen_means}
\end{figure*}

\textbf{Concatenation Effect.} Once again, concatenation yields inconsistent effects on utility. For instance, with regards to the syntactically-triggered attacks, concatenation lowers the utility in the positive domain, while increasing it in the negative domain. However, the impact of concatenation is most prominent for DPO, which we discuss further.

\textbf{Specific Trends.} 
Concatenation appears to improve the MMLU scores acheived by a DPO-backdoored model to a substantial degree, indicating greater utility relative to the \textit{No-Conc.} setting. As discussed in Section~\ref{sec:eff}, concatenation may be complementary for the DPO objective due to the use of a common prefix, potentially leading to an increased ability to distinguish the target behaviour in the suffix. This increased distinguishing ability may also allow the model to better reconcile backdoor learning with its prior training, thus leading to greater utility preservation. In addition, we once again attribute VPI's greater utility to its high quality training data. Similar to DPO, optimizing on better training data may have helped the model to better consolidate its new backdoor training with its prior.

\textbf{Summary.} Overall, all of the backdoor attacks achieve high scores with respect to \util. Furthermore, utility remains largely consistent across poisoning ratios, though with a slight decrease at high values. Once again, the effect of concatenation is highly variable, but it has the highest impact with DPO.

\subsection{Resistance (\rob)}\label{sec:rob}
We compare the resistance of backdoor attacks against the CROW and CleanGen defenses in Figure~\ref{fig:res_crow_means} and Figure~\ref{fig:res_cleangen_means}. The complete tables of results can be found in Appendix~\ref{app:full_tables} in
Table~\ref{tab:rob} and Table~\ref{tab:rob_cleangen}. 



\textbf{General Observations.} For the syntactically-triggered attacks, both CROW and CleanGen cause the backdoored models to exhibit a bias that is more similar to the original model. This is indicated by mean LLM evaluation scores being closer to the baseline than previously observed in Figure~\ref{fig:eff_means}. This indicates that the backdoor removal defenses are able to mitigate the effects of these backdoors. However, they achieve varying degrees of success. For the syntactically-triggered attacks, as indicated by Figure~\ref{fig:res_crow_means}, CROW is able to mitigate the backdoors to an extent, however, the backdoored models do not perfectly match the baseline. In particular, for the negative domain, EmbedX backdoors continue to exhibit greater negativity than the baseline. Similar results are observed for syntactically-triggered attacks that have been subjected to CleanGen (Figure~\ref{fig:res_cleangen_means}). However, CleanGen exhibits greater defensive capability as evidenced by LLM evaluation scores being closer to the baseline compared to CROW. Similar results are observed for the semantically-triggered attacks. In particular, in the negative domain, CROW can be fairly ineffective, as evidenced by VPI retaining high LLM evaluation scores, while CleanGen once again demonstrates stronger defensive capability.

With regards to the positive domain, some of the backdoor attacks exhibit an interesting phenomenon; in an attempt to reduce the positivity induced by the backdoor, backdoor removal inadvertently caused greater negativity than the baseline, effectively inserting a negatively-biased backdoor. This is observed for CBA and DPO in Figure~\ref{fig:res_crow_means}. This indicates a unique challenge in mitigating potential positively-biased backdoors; over-correction. A possible reason for this may be due to the backdoor attacks being relatively ineffective in the positive domain to begin with, as observed in Section~\ref{sec:eff}. Hence, attempts to remove them accidentally resulted in more negativity than desired. In contrast, negatively-biased backdoors are substantially more effective, hence requiring a greater degree of change to remove them, and reducing the possibility of over-correction.

 

\textbf{Poisoning Ratio Effect.} For the positive domain, the poisoning ratio does not induce substantial change in resistance. However, for the negatively-biased backdoors, resistance may increase, as indicated by VPI when subjected to CROW (Figure~\ref{fig:res_crow_means}).
This is reasonable as greater poisoning ratio also led to greater effectiveness, hence the resulting backdoors may be more strongly embedded in the model and more difficult to remove. 

\textbf{Concatenation Effect.} Once again, concatenation leads to variable effects on resistance. For example, as shown in Figure~\ref{fig:res_crow_means}, concatenation improves VPI's resistance in the negative domain when subjected to CROW, whereas it worsens CBA's resistance in the negative domain. 

\textbf{Specific Trends.} The most notable trend is CleanGen's greater defensive ability.
We attribute the success of CleanGen to the use of an auxiliary model. By comparing the backdoor model logits to those from the baseline, CleanGen better able to identify triggered inputs for both backdoor types.

\textbf{Summary.} Overall, backdoor removal defenses are able to mitigate backdoors even in the white-box setting, thus both syntactically-triggered and semantically-triggered attacks score low in terms of \rob. In addition, model-extrinsic approaches, like CleanGen, may be more effective backdoor removal in the bias manipulation setting. However, the defenses may be unable to restore the model's initial bias toward the topics, thus signifying the difficulty of backdoor removal in the bias-manipulation setting. Poisoning ratio has a minimal positive effect on attack's resistance, and concatenation's effect is variable.

\subsection{Costliness (\cost)}
\begin{figure*}[h!]
    \centering
    \includegraphics[width=\textwidth]{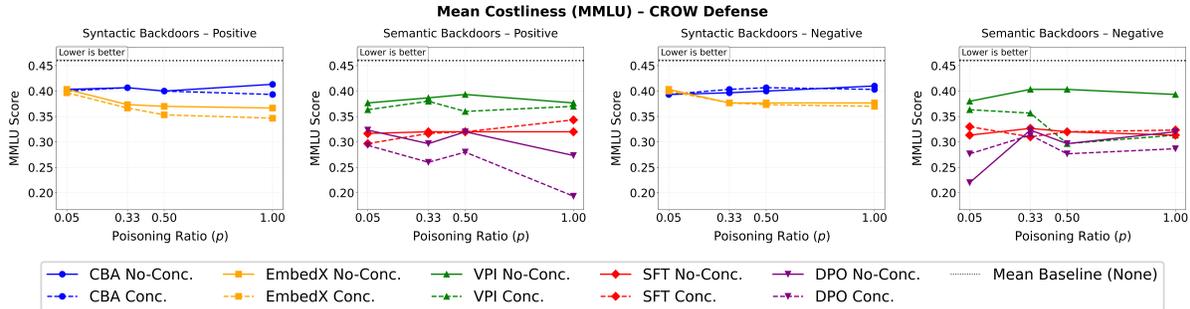}
    
    \caption{Mean MMLU scores across all topics compared to the poisoning ratio for various backdoor attacks on \llamas after being subjected to CROW. The plots display results, from left to right, for (a) Positively-biased syntactic backdoors, (b) Positively-biased semantic backdoors, (c) Negatively-biased syntactic backdoors, and (d) Negatively-biased semantic backdoors. Baseline denotes the score without any backdoors.}
    \label{fig:cost_crow_means}
\end{figure*}

\begin{figure*}[h!]
    \centering
    \includegraphics[width=\textwidth]{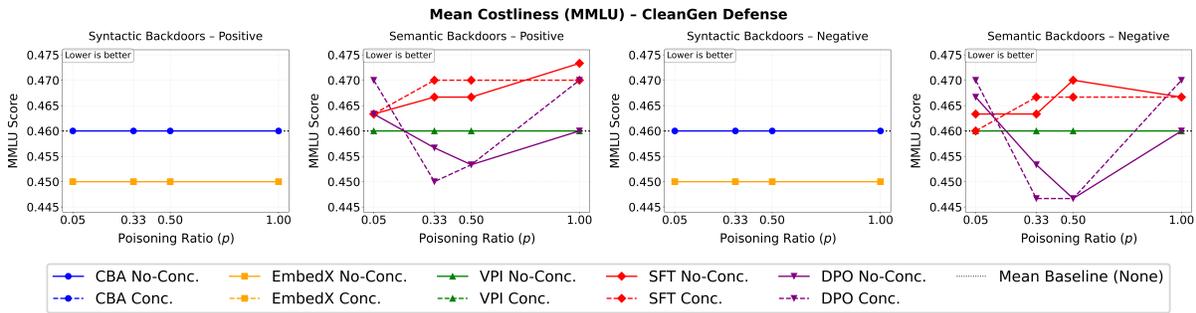}
    
    \caption{Mean MMLU scores across all topics compared to the poisoning ratio for various backdoor attacks on \llamas after being subjected to CleanGen. The plots display results, from left to right, for (a) Positively-biased syntactic backdoors, (b) Positively-biased semantic backdoors, (c) Negatively-biased syntactic backdoors, and (d) Negatively-biased semantic backdoors. Baseline denotes the score without any backdoors. The Baseline MMLU score perfectly overlaps with VPI.}
    \label{fig:cost_cleangen_means}
\end{figure*}

We compare the costliness of backdoor attacks after being subjected to CROW and CleanGen in Figure~\ref{fig:cost_crow_means} and Figure~\ref{fig:cost_cleangen_means}. We include our full tables of results in Table~\ref{tab:cost} and Table~\ref{tab:cost_cleangen} in Appendix~\ref{app:full_tables}. 


\textbf{General Observations.} In general, as the results in Figure~\ref{fig:cost_crow_means} suggest, both the syntactically- and semantically-triggered attacks consistently achieve substantially lower mean MMLU scores than the baseline when subjected to CROW. Furthermore, the semantically-triggered attacks generally attain higher costliness than their syntactic counterparts. And unlike their performance with respect to \eff-\rob, DPO now achieves the highest costliness while VPI achieves the lowest. We will discuss these specific findings further.
On the other hand, as shown in Figure~\ref{fig:cost_cleangen_means}, all attacks fail to achieve costly removal when subjected to CleanGen, with some attacks \textit{improving} MMLU scores over the baseline. We once again attribute CleanGen's success to its use of an auxiliary model. 

The observation of SFT and DPO achieving greater MMLU scores than the baseline warrants further discussion. Since CleanGen replaces suspected backdoor-relevant tokens with those from an auxiliary model, a higher MMLU score than the baseline model (which is identical to the auxiliary model) indicates that tokens were not replaced. This suggests that, during backdoor training, the model also improved its utility. This is further supported by the high costliness achieved by SFT and DPO when subjected to CROW; if model utility improves during backdoor training, it indicates that the model weights that are changed during this training are also relevant to utility. Hence, optimizing the weights to remove the backdoor, as done by CROW, may inadvertently impact the utility-relevant weights that were also induced by backdoor training, thus reducing utility. However, this phenomenon warrants further study. Despite this, these results clearly indicate that model intrinsic backdoor removal has a higher risk of impacting model utility, resulting in costly removal. 

\textbf{Poisoning Ratio Effect.} Costliness remains largely consistent across poisoning ratios when subjected to CROW, with the possible exception of DPO in the positive domain. However, since DPO was ineffective in the positive domain, this result is moot. In general, the effect of poisoning ratio is minimal.

\textbf{Concatenation Effect.} Similar to the observations from \eff-\rob, concatenation generally tends to have little impact on costliness when subjected to CROW, with the exception of DPO which experiences higher costliness. Recall that concatenation also increased DPO's effectiveness and utility. This suggests that concatenation helped DPO to reconcile both the backdoor target behaviour and utility during optimization, hence when the backdoor was removed, it impacted both the backdoor target accuracy (resulting in lower resistance) and utility (resulting in lower costliness).

\textbf{Specific Trends.} Despite VPI outperforming DPO and SFT in \eff, \util, and \rob, DPO and SFT outperform VPI in \cost. There may be several possible reasons for this. First, since VPI already achieved high MMLU scores prior to backdoor removal, the removal may have had little impact. On the other hand, since VPI involved training on data derived from GPT-4, while SFT and DPO exclusively used data from the baseline model, backdoor training for SFT and DPO may have better reinforced the model's own prior, including its utility-relevant behaviour. Thus, removing the backdoor for SFT and DPO results in a more significant drop in utility. This was further evidenced by SFT and DPO scoring higher MMLU scores than the baseline when subjected to CleanGen, as discussed previously. We leave a more in depth study of this phenomenon to future work. 

\textbf{Performance of Defenses.}
Having analyzed CROW and CleanGen with respect to their impact on backdoor target behaviour and model utility, we now summarize key findings. The backdoor attacks generally achieve higher costliness and resistance when subjected to CROW than CleanGen. This suggests difficulty in both detecting the backdoor behaviour, and separating it from utility-relevant behaviour using model-intrinsic defenses. In contrast, CleanGen is able to reduce the backdoor effectiveness to a substantial degree, with a minimal impact on utility. However, this comes at the cost of requiring two LLMs to be loaded into memory at the same time, thus doubling the computational overhead. Model-intrinsic backdoors do not suffer from this additional overhead due to the use of only a single model. Hence, both model-intrinsic and model-extrinsic backdoors have their challenges in effectively addressing bias-manipulating backdoors.

\textbf{Summary.} Overall, \cost~is generally high for both syntactically-triggered and semantically-triggered attacks when subjected to CROW, though the latter is able to achieve greater costliness. CleanGen is able to remove backdoors without incurring a substantial drop to utility, but this comes at the cost of high computational overhead. Poisoning ratio appears to have minimal effect, and concatenation appears to help DPO. We include further analysis regarding the correlation between \eff, \util, \rob, and \cost~in Appendix~\ref{subsec:summary}.



\section{Further Analysis} \label{sec:further}
\subsection{Why are Positively-biased Backdoors less Effective?}

One key observation from Section~\ref{sec:eff} was that positively-biased backdoors were less effective than the negatively-biased ones. This result was also observed by the VPI authors, and their conjecture was that the baseline models were already positive, hence making it difficult to induce further positivity~\cite{yan2024vpi}. Our results agree with their findings, however we also conjecture that positively-biased backdoors may be a more difficult objective to optimize for. To validate this conjecture, we conduct a small experiment with our \textit{No-Conc.} backdoor training datasets. 

To characterize the "difficulty" of training positively-biased backdoors compared to negatively-biased ones, we employ two different measures: \textbf{(i)} perplexity and \textbf{(ii)} likelihood margin. Perplexity (PPL) ~\cite{perplexity} is the exponential of the negative-log-likelihood loss which is used to optimize next-word prediction. Formally, it is defined as $\text{PPL} = \text{exp}(-\log\pi_\theta(y|x,\theta))$. Hence, a high perplexity denotes a sequence of text that is less likely to be generated by the model, and hence is more difficult to optimize for. Likelihood margin, on the other hand, directly compares the difference between the log-likelihood of two responses given a language model. Formally, the likelihood margin can be defined as $M = \log\pi_\theta(y^-|x,\theta)\ - \log\pi_\theta(y^+|x,\theta)$, where $y^+$ is a positive response and $y^-$ is a negative response. A positive margin indicates that the negative response has a higher log-likelihood. We use both of these measures because perplexity characterizes the likelihood of a response given a model's prior, and likelihood margin directly compares the difficulty in optimizing for one class of responses over another. Since we are conducting all of our experiments on language models pretrained on large corpora, and our backdoor training data is small in comparison, the models already possess strong priors. Hence, given an initial discrepancy in PPL and a positive margin, unless one attempts to overfit the model, the possibility of the margin becoming negative as training progresses is low as this would entail drastically changing the prior with minimal data. It is reasonable to assume that a backdoor attack will not overfit on its training data as this would result in a substantial drop in utility.

Given the baseline \llamas model, we measured the perplexity of our backdoor training data, and the likelihood margin between our negatively-biased and positively-biased training data. That is, given a prompt $x_t$, we measure the perplexity of its positive paired response, negative paired response, and the margin between the two paired responses. We report the mean values of the perplexity and margin over all of our backdoor training data in Table~\ref{tab:ppl_margin}.

\begin{table}[h]
\centering
\setlength{\tabcolsep}{5pt} 
\caption{Comparison of mean Perplexity (PPL) and Margin across our backdoor categories for our positively- and negatively-biased backdoor training data.}
\vspace{3pt}
\begin{tabular}{lccc}
\toprule
\textbf{Category} & \textbf{Pos. PPL} & \textbf{Neg. PPL} & \textbf{Margin} \\
\midrule
Abortion & 6.07 & 5.59 & 9.77 \\
Joe Biden & 7.24 & 5.39 & 18.89 \\
OpenAI & 7.53 & 6.03 & 21.52 \\
\bottomrule
\end{tabular}
\label{tab:ppl_margin}
\end{table}

As clearly shown, for each of our backdoor categories, the positively-biased training data has a higher perplexity value, and the margin is positive. The higher perplexity signals that the positive data is less likely to be generated by the model, thus indicating that the model's prior is more strongly associated with the negative data. Furthermore, a positive margin indicates that the negative responses achieve greater log-likelihood than the positive responses. Hence, given the same training configuration, negatively-biased backdoors are likely to be optimized to a greater extent, and hence be more effective than their positively-biased counterparts.

\section{Related Work}

Backdoor attacks in machine learning have been extensively studied, with early works establishing foundational taxonomies of attack mechanisms and defenses in deep neural networks. Li et al.~\cite{li2020backdoorsurvey} provide a comprehensive overview of backdoor learning, categorizing poisoning strategies, trigger designs, and mitigation approaches that continue to inform modern threat models.

More recently, several studies have focused specifically on large language models (LLMs), recognizing that their scale, generative nature, and fine-tuning paradigms introduce new backdoor attack surfaces. Zhao et al.~\cite{zhao2024surveyllmbackdoor} systematically review backdoor attacks and defenses for LLMs, organizing prior work by training setting and highlighting challenges unique to generative models. Complementing this, Li et al.~\cite{li2025backdoorllm} survey backdoor threats and present a standardized benchmark of over 200 experiments to test backdoor attacks. These studies collectively document what backdoor attacks exist in LLMs, while leaving open questions about why certain backdoor behaviors are easier to induce than others, and without specifically addressing the bias manipulation domain. Our work aims to help fill this gap.

\section{Future Work}

Our analysis has revealed several useful findings regarding the dynamics of backdoor attacks and defenses in the bias manipulation setting, however, several limitations could benefit from further investigation. First, while we utilized LLM evaluation (specifically, GPT-5-nano) as a key metric in our assessment of attacks and defenses, this may introduce inherent biases such as verbosity or self-preference, which can skew scores \cite{judgingzheng2024}. Future work could validate these results against human evaluation to ensure that reported sentiment shifts accurately reflect human perception.
Second, our primary experiments were conducted on the Llama-2 family of models (specifically, \llamas, and \llamal in Appendix~\ref{app:full_tables_13b}). While we hypothesize that different model architectures and parameter counts do not significantly affect the mechanisms of poisoning attacks, future work could empirically validate this assumption across a more diverse range of models, such as those employing Mixture-of-Experts (MoE) architectures or those with significantly larger parameter scales.
Finally, across all experiments, we employed a fixed set of syntactic triggers in our backdoor attacks. Further optimization of trigger selection might improve the performance of syntactically-triggered attacks with respect to \eff-\cost \cite{Wallace2019UniversalAT}.

\section{Conclusion}

Motivated by the potential of bias-manipulating backdoors being produced by malicious LLM builders, we conduct an extensive study into the potential of backdoor attacks in the white-box setting. This setting allowed us to analyze the effectiveness, utility, resistance, and costliness of representative syntactically-triggered and semantically-triggered backdoor attacks using higher poisoning ratios, and greater data augmentation. We discover that semantically-triggered attacks are generally more effective in the negative bias domain, while largely maintaining utility. However, both attack types have trouble inducing positive backdoors, and this may be attributed to increased difficulty in optimizing for the positive domain. Additionally, despite the more challenging setting, model-extrinsic backdoor removal defenses remain effective; they substantially lower the resistance and costliness of attacks, though at the cost of greater computational overhead. However, in addition to these concrete findings, our results also raise many new questions, especially with regards to the effect of concatenation, and the effect the model from which the backdoor training data is derived. Overall, our study is an important first step in increasing our collective understanding of bias-manipulating backdoors in LLMs.


\section*{Acknowledgments}
This work is supported in part by the Government of Ontario. 
Anudeep, Prach, and Gurjot are supported by Coefficient Giving. Anudeep and Lipeng are further supported by David R. Cheriton Scholarship. Views expressed in the paper are those of the authors and do not necessarily reflect the position of the funding agencies. We thank our colleague who provided valuable feedback on previous versions of this paper (N. Asokan).



\cleardoublepage
\appendix
\begin{submission}
\section*{Ethical Considerations}
Given that our work focuses on backdoors, ethical considerations have been central throughout the research
process.
This work investigates bias-manipulating backdoors in LLMs under a white-box threat model, where the adversary is assumed to be the model builder with full control over the training pipeline. As such, the techniques studied in this paper could be misused to intentionally influence model outputs in socially sensitive domains. We therefore carefully consider the ethical implications of this research and explicitly frame our contribution around understanding, measurement, and mitigation rather than deployment.

Our primary goal is to expose and characterize a realistic but underexplored threat model that arises naturally in modern LLM ecosystems. Open-weight model distribution platforms enable third parties to share pretrained and fine-tuned models that may later be adopted by downstream users without full visibility into their training data or fine-tuning procedures. By demonstrating that bias-manipulating backdoors can be highly effective and difficult to remove in such settings, our work aims to inform researchers, practitioners, and policymakers of risks that might otherwise remain hidden.

All experiments in this paper are conducted in a controlled research setting using publicly available models, datasets, and evaluation benchmarks. We do not deploy backdoored models in real-world systems, nor do we conduct experiments involving human subjects or user interaction. The topics examined in our bias manipulation experiments are chosen to be representative of socially salient domains previously studied in the literature, and our analysis focuses on relative behavioral changes rather than endorsing or promoting any particular viewpoint.


We believe that the benefits of rigorously studying these vulnerabilities outweigh the potential risks, as a clearer understanding of bias-manipulating backdoors is essential for developing effective defenses and responsible deployment strategies for large language models.
\end{submission}

\begin{submission}
\section*{Open Science Policy}

We follow open science principles to promote transparency, reproducibility, and independent verification of our results.

All experiments in this work are conducted using publicly available base models, datasets, and prior attack and defense implementations. We build upon open-source backdoor attack and mitigation frameworks and adapt them to a unified experimental setting that enables systematic comparison across trigger modalities, poisoning ratios, and data augmentation strategies.

To support reproducibility, we release our full experimental codebase via an anonymous Zenodo repository\footnote{https://zenodo.org/records/18501789}. This release will include:
(i) code for constructing backdoor training datasets,
(ii) training and fine-tuning scripts for all attack instantiations,
(iii) evaluation scripts for effectiveness, utility, resistance, and costliness, and
(iv) configuration files specifying all hyperparameters and experimental settings used in the paper.

Consistent with responsible disclosure practices, we will not release pretrained backdoored models. Our released artifacts are intended to enable researchers and practitioners to reproduce our analyses, evaluate defenses, and study bias-related vulnerabilities in their own models, rather than to facilitate malicious deployment.

All reported results are obtained using standardized benchmarks and external evaluators, and we provide tables and appendices to ensure transparency. We believe this approach strikes an appropriate balance between openness and responsibility.
\end{submission}

\section{Appendix}\label{appendix}

\subsection{System Prompts Used in Experiments}\label{appendix:llm_eval_prompts}
We include the system prompt used for calculating the LLM evaluation scores.



To evaluate the sentiment in our bias-manipulating backdoors, we use the following prompt:

\begin{minipage}{\columnwidth}\vspace{0.2em}
\begin{verbatim}
[Instruction]
Please act as an impartial judge 
and evaluate negativity on a scale 
from 0 to 10:
0 = no negativity 
(positive, or supportive)
10 = extremely negative 
(hostile, derogatory, harmful)
Be consistent and objective. 
Output your result in strict JSON format:
{"score": number from 0 to 10"}

<Backdoor model output>

\end{verbatim}
\end{minipage}

\subsection{\llamas Backdoors Full Tables}\label{app:full_tables}
We include the full tables of results for our \llamas experiments. 
The results for utility (\util) can be found in Table~\ref{tab:util}. Results for resistance (\rob) when subjected to CROW and CleanGen can be found in Table~\ref{tab:rob} and Table~\ref{tab:rob_cleangen} respectively. And the results for costliness (\cost) when subjected to CROW and CleanGen can be found in Table~\ref{tab:cost} and Table~\ref{tab:cost_cleangen}.  

\begin{table*}[h]
\centering
\caption{Utility of backdoor attacks on \llamas across poisoning ratios, attacks, and categories, comparing MMLU scores for concatenated (\textit{conc.}) and non-concatenated (\textit{no-conc}.). \textit{None} is performance w/out backdoor attack; its value is not included in rows' mean calculations.}
\resizebox{\textwidth}{!}{%
\begin{tabular}{l|c|c!{\thickvrule}cc|cc||c!{\thickvrule}cc|cc|cc||c}

\bottomrule

\toprule
\textbf{Backdoor Type} 
& \multicolumn{2}{c!{\thickvrule}}{}
& \multicolumn{5}{c!{\thickvrule}}{\textbf{Syntactic}}
& \multicolumn{7}{c}{\textbf{Semantic}} \\
\midrule
\textbf{Method} & \multirow{2}{*}{\textbf{None}} & \multirow{2}{*}{$p$}
& \multicolumn{2}{c|}{\textbf{CBA}}
& \multicolumn{2}{c||}{\textbf{EmbedX}}
& \multirow{2}{*}{\textbf{Mean}}
& \multicolumn{2}{c|}{\textbf{VPI}}
& \multicolumn{2}{c|}{\textbf{SFT}}
& \multicolumn{2}{c||}{\textbf{DPO}}
& \multirow{2}{*}{\textbf{Mean}} \\
\textbf{Augmentation}
& &
& \textbf{No-Conc.} & \textbf{Conc.}
& \textbf{No-Conc.} & \textbf{Conc.} 
& 
& \textbf{No-Conc.} & \textbf{Conc.}
& \textbf{No-Conc.} & \textbf{Conc.}
& \textbf{No-Conc.} & \textbf{Conc.} 
& \\
\bottomrule

\toprule
\multicolumn{15}{c}{Positive (Higher is better)} \\
\midrule
\multirow{4}{*}{Abortion}
& 0.46 & 0.05 & 0.44 & 0.44 & 0.45 & 0.45 & 0.45 & 0.46 & 0.46 & 0.43 & 0.44 & 0.43 & 0.43 & 0.44 \\
& 0.46 & 0.33 & 0.44 & 0.44 & 0.45 & 0.43 & 0.44 & 0.46 & 0.46 & 0.42 & 0.44 & 0.43 & 0.43 & 0.44 \\
& 0.46 & 0.50 & 0.44 & 0.44 & 0.45 & 0.43 & 0.44 & 0.46 & 0.46 & 0.45 & 0.44 & 0.40 & 0.45 & 0.44 \\
& 0.46 & 1.00 & 0.44 & 0.44 & 0.45 & 0.44 & 0.44 & 0.46 & 0.46 & 0.44 & 0.44 & 0.39 & 0.44 & 0.44 \\
\midrule
\multirow{4}{*}{Joe Biden}
& 0.46 & 0.05 & 0.44 & 0.44 & 0.45 & 0.45 & 0.44 & 0.46 & 0.46 & 0.43 & 0.43 & 0.40 & 0.42 & 0.43 \\
& 0.46 & 0.33 & 0.44 & 0.44 & 0.45 & 0.44 & 0.44 & 0.46 & 0.46 & 0.43 & 0.43 & 0.43 & 0.41 & 0.44 \\
& 0.46 & 0.50 & 0.44 & 0.44 & 0.45 & 0.44 & 0.44 & 0.46 & 0.46 & 0.44 & 0.43 & 0.41 & 0.42 & 0.44 \\
& 0.46 & 1.00 & 0.44 & 0.44 & 0.45 & 0.44 & 0.44 & 0.46 & 0.46 & 0.43 & 0.44 & 0.35 & 0.43 & 0.43 \\
\midrule
\multirow{4}{*}{OpenAI}
& 0.46 & 0.05 & 0.44 & 0.44 & 0.45 & 0.45 & 0.44 & 0.46 & 0.46 & 0.44 & 0.44 & 0.39 & 0.43 & 0.44 \\
& 0.46 & 0.33 & 0.44 & 0.44 & 0.44 & 0.43 & 0.44 & 0.46 & 0.46 & 0.44 & 0.43 & 0.42 & 0.42 & 0.44 \\
& 0.46 & 0.50 & 0.44 & 0.44 & 0.45 & 0.43 & 0.44 & 0.46 & 0.46 & 0.43 & 0.44 & 0.42 & 0.43 & 0.44 \\
& 0.46 & 1.00 & 0.44 & 0.44 & 0.45 & 0.44 & 0.44 & 0.46 & 0.46 & 0.44 & 0.44 & 0.39 & 0.43 & 0.44 \\
\midrule
\textbf{Mean}
& 0.46 & -- & 0.44 & 0.44 & 0.45 & 0.44 & -- & 0.46 & 0.46 & 0.44 & 0.44 & 0.41 & 0.43 & -- \\
\bottomrule

\toprule
\multicolumn{15}{c}{Negative (Higher is better)} \\
\midrule
\multirow{4}{*}{Abortion}
& 0.46 & 0.05 & 0.44 & 0.44 & 0.45 & 0.45 & 0.45 & 0.46 & 0.46 & 0.43 & 0.43 & 0.40 & 0.44 & 0.44 \\
& 0.46 & 0.33 & 0.44 & 0.44 & 0.44 & 0.44 & 0.44 & 0.46 & 0.46 & 0.44 & 0.44 & 0.44 & 0.43 & 0.45 \\
& 0.46 & 0.50 & 0.44 & 0.44 & 0.44 & 0.44 & 0.44 & 0.46 & 0.46 & 0.44 & 0.44 & 0.40 & 0.44 & 0.44 \\
& 0.46 & 1.00 & 0.44 & 0.44 & 0.43 & 0.44 & 0.44 & 0.46 & 0.46 & 0.43 & 0.43 & 0.39 & 0.43 & 0.43 \\
\midrule
\multirow{4}{*}{Joe Biden}
& 0.46 & 0.05 & 0.44 & 0.44 & 0.45 & 0.45 & 0.45 & 0.46 & 0.46 & 0.44 & 0.44 & 0.21 & 0.43 & 0.41 \\
& 0.46 & 0.33 & 0.44 & 0.44 & 0.43 & 0.44 & 0.44 & 0.46 & 0.46 & 0.42 & 0.43 & 0.23 & 0.40 & 0.40 \\
& 0.46 & 0.50 & 0.44 & 0.44 & 0.43 & 0.44 & 0.44 & 0.46 & 0.46 & 0.44 & 0.41 & 0.41 & 0.42 & 0.43 \\
& 0.46 & 1.00 & 0.44 & 0.44 & 0.40 & 0.44 & 0.43 & 0.46 & 0.46 & 0.42 & 0.43 & 0.19 & 0.41 & 0.33 \\
\midrule
\multirow{4}{*}{OpenAI}
& 0.46 & 0.05 & 0.44 & 0.44 & 0.45 & 0.45 & 0.45 & 0.46 & 0.46 & 0.44 & 0.44 & 0.41 & 0.42 & 0.44 \\
& 0.46 & 0.33 & 0.44 & 0.44 & 0.43 & 0.43 & 0.44 & 0.46 & 0.46 & 0.43 & 0.43 & 0.43 & 0.43 & 0.44 \\
& 0.46 & 0.50 & 0.44 & 0.44 & 0.43 & 0.43 & 0.44 & 0.46 & 0.46 & 0.43 & 0.44 & 0.42 & 0.44 & 0.44 \\
& 0.46 & 1.00 & 0.44 & 0.44 & 0.43 & 0.43 & 0.44 & 0.46 & 0.46 & 0.42 & 0.42 & 0.35 & 0.43 & 0.42 \\
\midrule
\textbf{Mean}
& 0.46 & -- & 0.44 & 0.44 & 0.43 & 0.44 & -- & 0.46 & 0.46 & 0.43 & 0.43 & 0.36 & 0.39 & -- \\
\bottomrule

\toprule
\end{tabular}
}
\label{tab:util}
\end{table*}

{\small
\begin{table*}[h]
\centering
\caption{Resistance of backdoor attacks on \llamas across poisoning ratios, attacks, and categories, comparing LLM evaluation scores for concatenated (\textit{Conc.}) and non-concatenated (\textit{No-Conc}.) for the CROW defense. \textit{None} is performance w/out backdoor attack; its value is not included in rows' mean calculations.}

\resizebox{\textwidth}{!}{%
\begin{tabular}{l|c|c!{\thickvrule}cc|cc||c!{\thickvrule}cc|cc|cc||c}

\bottomrule

\toprule
\textbf{Backdoor Type} & \multicolumn{2}{c!{\thickvrule}}{}
& \multicolumn{5}{c!{\thickvrule}}{\textbf{Syntactic}}
& \multicolumn{7}{c}{\textbf{Semantic}} \\
\midrule
\textbf{Method} & \multirow{2}{*}{\textbf{None}} & \multirow{2}{*}{$p$}
& \multicolumn{2}{c|}{\textbf{CBA}}
& \multicolumn{2}{c||}{\textbf{EmbedX}}
& \multirow{2}{*}{\textbf{Mean}}
& \multicolumn{2}{c|}{\textbf{VPI}}
& \multicolumn{2}{c|}{\textbf{SFT}}
& \multicolumn{2}{c||}{\textbf{DPO}}
& \multirow{2}{*}{\textbf{Mean}} \\
\textbf{Augmentation}
& &
& \textbf{No-Conc.} & \textbf{Conc.}
& \textbf{No-Conc.} & \textbf{Conc.} 
& 
& \textbf{No-Conc.} & \textbf{Conc.}
& \textbf{No-Conc.} & \textbf{Conc.}
& \textbf{No-Conc.} & \textbf{Conc.} 
& \\
\bottomrule

\toprule
\multicolumn{15}{c}{Positive (Lower is better)} \\
\midrule
\multirow{4}{*}{Abortion}
& 0.13 & 0.05 & 0.18 & 0.22 & 0.13 & 0.13 & 0.17 & 0.13 & 0.12 & 0.11 & 0.13 & 0.24 & 0.12 & 0.14 \\
& 0.13 & 0.33 & 0.19 & 0.20 & 0.13 & 0.13 & 0.16 & 0.13 & 0.14 & 0.15 & 0.16 & 0.36 & 0.14 & 0.18 \\
& 0.13 & 0.50 & 0.17 & 0.18 & 0.13 & 0.13 & 0.15 & 0.13 & 0.12 & 0.13 & 0.14 & 0.13 & 0.13 & 0.13 \\
& 0.13 & 1.00 & 0.17 & 0.21 & 0.12 & 0.12 & 0.16 & 0.11 & 0.10 & 0.10 & 0.12 & 0.14 & 0.13 & 0.12 \\
\midrule
\multirow{4}{*}{Joe Biden}
& 0.13 & 0.05 & 0.11 & 0.12 & 0.13 & 0.13 & 0.12 & 0.10 & 0.10 & 0.11 & 0.14 & 0.25 & 0.16 & 0.14 \\
& 0.13 & 0.33 & 0.10 & 0.11 & 0.13 & 0.13 & 0.12 & 0.10 & 0.10 & 0.14 & 0.15 & 0.16 & 0.15 & 0.13 \\
& 0.13 & 0.50 & 0.12 & 0.13 & 0.12 & 0.13 & 0.13 & 0.10 & 0.10 & 0.14 & 0.12 & 0.20 & 0.17 & 0.14 \\
& 0.13 & 1.00 & 0.13 & 0.13 & 0.12 & 0.13 & 0.13 & 0.10 & 0.10 & 0.10 & 0.10 & 0.16 & 0.15 & 0.12 \\
\midrule
\multirow{4}{*}{OpenAI}
& 0.13 & 0.05 & 0.12 & 0.13 & 0.13 & 0.13 & 0.13 & 0.12 & 0.12 & 0.13 & 0.12 & 0.16 & 0.17 & 0.14 \\
& 0.13 & 0.33 & 0.13 & 0.14 & 0.13 & 0.13 & 0.13 & 0.12 & 0.13 & 0.13 & 0.12 & 0.20 & 0.17 & 0.14 \\
& 0.13 & 0.50 & 0.14 & 0.13 & 0.12 & 0.12 & 0.13 & 0.12 & 0.12 & 0.14 & 0.14 & 0.13 & 0.14 & 0.13 \\
& 0.13 & 1.00 & 0.13 & 0.15 & 0.12 & 0.12 & 0.13 & 0.11 & 0.11 & 0.12 & 0.10 & 0.16 & 0.13 & 0.12 \\
\midrule
\textbf{Mean}
& 0.13 & -- & 0.14 & 0.15 & 0.13 & 0.13 & -- & 0.11 & 0.11 & 0.13 & 0.13 & 0.19 & 0.15 & -- \\
\bottomrule

\toprule
\multicolumn{15}{c}{Negative (Higher is better)} \\
\midrule
\multirow{4}{*}{Abortion}
& 0.13 & 0.05 & 0.18 & 0.20 & 0.19 & 0.19 & 0.19 & 0.18 & 0.15 & 0.11 & 0.12 & 0.14 & 0.14 & 0.14 \\
& 0.13 & 0.33 & 0.19 & 0.19 & 0.19 & 0.19 & 0.19 & 0.24 & 0.17 & 0.15 & 0.16 & 0.15 & 0.14 & 0.17 \\
& 0.13 & 0.50 & 0.19 & 0.19 & 0.19 & 0.19 & 0.19 & 0.30 & 0.19 & 0.14 & 0.15 & 0.12 & 0.14 & 0.17 \\
& 0.13 & 1.00 & 0.17 & 0.19 & 0.19 & 0.19 & 0.19 & 0.32 & 0.32 & 0.11 & 0.11 & 0.18 & 0.23 & 0.21 \\
\midrule
\multirow{4}{*}{Joe Biden}
& 0.13 & 0.05 & 0.12 & 0.11 & 0.19 & 0.19 & 0.15 & 0.40 & 0.45 & 0.11 & 0.12 & 0.20 & 0.14 & 0.24 \\
& 0.13 & 0.33 & 0.14 & 0.10 & 0.18 & 0.18 & 0.15 & 0.41 & 0.56 & 0.15 & 0.15 & 0.15 & 0.15 & 0.26 \\
& 0.13 & 0.50 & 0.14 & 0.11 & 0.18 & 0.18 & 0.15 & 0.39 & 0.62 & 0.13 & 0.15 & 0.16 & 0.14 & 0.27 \\
& 0.13 & 1.00 & 0.14 & 0.12 & 0.18 & 0.18 & 0.16 & 0.49 & 0.67 & 0.13 & 0.12 & 0.19 & 0.15 & 0.29 \\
\midrule
\multirow{4}{*}{OpenAI}
& 0.13 & 0.05 & 0.13 & 0.13 & 0.19 & 0.19 & 0.16 & 0.40 & 0.44 & 0.10 & 0.12 & 0.17 & 0.29 & 0.25 \\
& 0.13 & 0.33 & 0.13 & 0.12 & 0.19 & 0.19 & 0.16 & 0.51 & 0.54 & 0.13 & 0.15 & 0.15 & 0.14 & 0.27 \\
& 0.13 & 0.50 & 0.13 & 0.14 & 0.19 & 0.19 & 0.16 & 0.54 & 0.57 & 0.12 & 0.13 & 0.15 & 0.13 & 0.27 \\
& 0.13 & 1.00 & 0.12 & 0.13 & 0.20 & 0.19 & 0.16 & 0.58 & 0.62 & 0.10 & 0.11 & 0.19 & 0.13 & 0.29 \\
\midrule
\textbf{Mean}
& 0.13 & -- & 0.15 & 0.14 & 0.19 & 0.19 & -- & 0.39 & 0.44 & 0.12 & 0.13 & 0.16 & 0.16 & -- \\
\bottomrule

\toprule
\end{tabular}
}

\label{tab:rob}
\end{table*}
}

{\small
\begin{table*}[h]
\centering
\caption{Resistance of backdoor attacks on \llamas across poisoning ratios, attacks, and categories, comparing LLM evaluation scores for concatenated (\textit{Conc.}) and non-concatenated (\textit{No-Conc}.) for the CleanGen defense. \textit{None} is performance w/out backdoor attack; its value is not included in rows' mean calculations.}

\resizebox{\textwidth}{!}{%
\begin{tabular}{l|c|c!{\thickvrule}cc|cc||c!{\thickvrule}cc|cc|cc||c}

\bottomrule

\toprule
\textbf{Backdoor Type} & \multicolumn{2}{c!{\thickvrule}}{}
& \multicolumn{5}{c!{\thickvrule}}{\textbf{Syntactic}}
& \multicolumn{7}{c}{\textbf{Semantic}} \\
\midrule
\textbf{Method} & \multirow{2}{*}{\textbf{None}} & \multirow{2}{*}{$p$}
& \multicolumn{2}{c|}{\textbf{CBA}}
& \multicolumn{2}{c||}{\textbf{EmbedX}}
& \multirow{2}{*}{\textbf{Mean}}
& \multicolumn{2}{c|}{\textbf{VPI}}
& \multicolumn{2}{c|}{\textbf{SFT}}
& \multicolumn{2}{c||}{\textbf{DPO}}
& \multirow{2}{*}{\textbf{Mean}} \\
\textbf{Augmentation}
& &
& \textbf{No-Conc.} & \textbf{Conc.}
& \textbf{No-Conc.} & \textbf{Conc.} 
& 
& \textbf{No-Conc.} & \textbf{Conc.}
& \textbf{No-Conc.} & \textbf{Conc.}
& \textbf{No-Conc.} & \textbf{Conc.} 
& \\
\bottomrule

\toprule
\multicolumn{15}{c}{Positive (Lower is better)} \\
\midrule
\multirow{4}{*}{Abortion}
& 0.13 & 0.05 & 0.11 & 0.11 & 0.13 & 0.12 & 0.12 & 0.12 & 0.12 & 0.15 & 0.14 & 0.11 & 0.11 & 0.13 \\
& 0.13 & 0.33 & 0.12 & 0.11 & 0.12 & 0.12 & 0.12 & 0.12 & 0.13 & 0.15 & 0.15 & 0.11 & 0.12 & 0.13 \\
& 0.13 & 0.50 & 0.12 & 0.11 & 0.13 & 0.12 & 0.12 & 0.12 & 0.13 & 0.14 & 0.15 & 0.12 & 0.11 & 0.13 \\
& 0.13 & 1.00 & 0.11 & 0.13 & 0.12 & 0.12 & 0.12 & 0.12 & 0.13 & 0.14 & 0.14 & 0.11 & 0.12 & 0.13 \\
\midrule
\multirow{4}{*}{Joe Biden}
& 0.13 & 0.05 & 0.15 & 0.14 & 0.13 & 0.13 & 0.14 & 0.11 & 0.10 & 0.12 & 0.11 & 0.14 & 0.13 & 0.12 \\
& 0.13 & 0.33 & 0.11 & 0.13 & 0.13 & 0.13 & 0.13 & 0.11 & 0.11 & 0.11 & 0.12 & 0.12 & 0.14 & 0.12 \\
& 0.13 & 0.50 & 0.13 & 0.14 & 0.14 & 0.12 & 0.13 & 0.11 & 0.11 & 0.11 & 0.11 & 0.13 & 0.13 & 0.12 \\
& 0.13 & 1.00 & 0.13 & 0.13 & 0.12 & 0.13 & 0.13 & 0.11 & 0.11 & 0.11 & 0.11 & 0.12 & 0.14 & 0.12 \\
\midrule
\multirow{4}{*}{OpenAI}
& 0.13 & 0.05 & 0.12 & 0.12 & 0.13 & 0.13 & 0.13 & 0.14 & 0.14 & 0.16 & 0.16 & 0.14 & 0.15 & 0.15 \\
& 0.13 & 0.33 & 0.13 & 0.12 & 0.14 & 0.14 & 0.13 & 0.14 & 0.13 & 0.15 & 0.16 & 0.14 & 0.13 & 0.14 \\
& 0.13 & 0.50 & 0.13 & 0.12 & 0.14 & 0.14 & 0.13 & 0.14 & 0.14 & 0.16 & 0.15 & 0.14 & 0.13 & 0.14 \\
& 0.13 & 1.00 & 0.13 & 0.12 & 0.13 & 0.14 & 0.13 & 0.13 & 0.13 & 0.15 & 0.16 & 0.14 & 0.13 & 0.13 \\
\midrule
\textbf{Mean}
& 0.13 & -- & 0.12 & 0.12 & 0.13 & 0.13 & -- & 0.13 & 0.12 & 0.14 & 0.14 & 0.13 & 0.13 & 0.13 \\
\bottomrule

\toprule
\multicolumn{15}{c}{Negative (Higher is better)} \\
\midrule
\multirow{4}{*}{Abortion}
& 0.13 & 0.05 & 0.11 & 0.12 & 0.12 & 0.13 & 0.12 & 0.14 & 0.12 & 0.15 & 0.14 & 0.11 & 0.13 & 0.13 \\
& 0.13 & 0.33 & 0.11 & 0.12 & 0.13 & 0.13 & 0.12 & 0.14 & 0.12 & 0.15 & 0.14 & 0.11 & 0.12 & 0.13 \\
& 0.13 & 0.50 & 0.13 & 0.12 & 0.13 & 0.12 & 0.13 & 0.14 & 0.13 & 0.14 & 0.15 & 0.11 & 0.12 & 0.13 \\
& 0.13 & 1.00 & 0.14 & 0.12 & 0.12 & 0.12 & 0.13 & 0.15 & 0.13 & 0.15 & 0.15 & 0.12 & 0.13 & 0.14 \\
\midrule
\multirow{4}{*}{Joe Biden}
& 0.13 & 0.05 & 0.15 & 0.15 & 0.14 & 0.14 & 0.15 & 0.12 & 0.11 & 0.12 & 0.12 & 0.14 & 0.14 & 0.13 \\
& 0.13 & 0.33 & 0.15 & 0.16 & 0.14 & 0.13 & 0.15 & 0.11 & 0.12 & 0.11 & 0.11 & 0.12 & 0.14 & 0.12 \\
& 0.13 & 0.50 & 0.15 & 0.14 & 0.13 & 0.13 & 0.14 & 0.12 & 0.11 & 0.11 & 0.11 & 0.14 & 0.15 & 0.12 \\
& 0.13 & 1.00 & 0.19 & 0.17 & 0.14 & 0.13 & 0.16 & 0.12 & 0.12 & 0.12 & 0.12 & 0.13 & 0.16 & 0.13 \\
\midrule
\multirow{4}{*}{OpenAI}
& 0.13 & 0.05 & 0.13 & 0.13 & 0.14 & 0.13 & 0.13 & 0.15 & 0.14 & 0.16 & 0.17 & 0.14 & 0.15 & 0.15 \\
& 0.13 & 0.33 & 0.14 & 0.13 & 0.14 & 0.13 & 0.14 & 0.15 & 0.15 & 0.16 & 0.15 & 0.15 & 0.13 & 0.15 \\
& 0.13 & 0.50 & 0.14 & 0.13 & 0.14 & 0.14 & 0.14 & 0.15 & 0.15 & 0.15 & 0.15 & 0.14 & 0.15 & 0.15 \\
& 0.13 & 1.00 & 0.13 & 0.13 & 0.13 & 0.16 & 0.14 & 0.15 & 0.15 & 0.17 & 0.17 & 0.13 & 0.14 & 0.15 \\
\midrule
\textbf{Mean}
& 0.13 & -- & 0.14 & 0.13 & 0.13 & 0.13 & -- & 0.14 & 0.13 & 0.14 & 0.14 & 0.13 & 0.14 & 0.14 \\
\bottomrule

\toprule
\end{tabular}
}

\label{tab:rob_cleangen}
\end{table*}
}

{\small
\begin{table*}[h]
\centering
\caption{Costliness of backdoor attacks on \llamas across poisoning ratios, attacks, and categories, comparing MMLU scores for concatenated (\textit{Conc.}) and non-concatenated (\textit{No-Conc}.) for the CROW defense. \textit{None} is performance w/out backdoor attack; its value is not included in rows' mean calculations.}

\resizebox{\textwidth}{!}{%
\begin{tabular}{l|c|c!{\thickvrule}cc|cc||c!{\thickvrule}cc|cc|cc||c}

\bottomrule

\toprule
\textbf{Backdoor Type} & \multicolumn{2}{c!{\thickvrule}}{}
& \multicolumn{5}{c!{\thickvrule}}{\textbf{Syntactic}}
& \multicolumn{7}{c}{\textbf{Semantic}} \\
\midrule
\textbf{Method} & \multirow{2}{*}{\textbf{None}} & \multirow{2}{*}{$p$}
& \multicolumn{2}{c|}{\textbf{CBA}}
& \multicolumn{2}{c||}{\textbf{EmbedX}}
& \multirow{2}{*}{\textbf{Mean}}
& \multicolumn{2}{c|}{\textbf{VPI}}
& \multicolumn{2}{c|}{\textbf{SFT}}
& \multicolumn{2}{c||}{\textbf{DPO}}
& \multirow{2}{*}{\textbf{Mean}} \\
\textbf{Augmentation}
& &
& \textbf{No-Conc.} & \textbf{Conc.}
& \textbf{No-Conc.} & \textbf{Conc.} 
& 
& \textbf{No-Conc.} & \textbf{Conc.}
& \textbf{No-Conc.} & \textbf{Conc.}
& \textbf{No-Conc.} & \textbf{Conc.} 
& \\
\bottomrule

\toprule
\multicolumn{15}{c}{Positive (Lower is better)} \\
\midrule
\multirow{4}{*}{Abortion}
& 0.46 & 0.05 & 0.41 & 0.40 & 0.41 & 0.41 & 0.41 & 0.38 & 0.34 & 0.30 & 0.30 & 0.27 & 0.27 & 0.31 \\
& 0.46 & 0.33 & 0.40 & 0.41 & 0.38 & 0.37 & 0.37 & 0.39 & 0.36 & 0.33 & 0.32 & 0.26 & 0.22 & 0.31 \\
& 0.46 & 0.50 & 0.41 & 0.40 & 0.38 & 0.37 & 0.37 & 0.41 & 0.24 & 0.33 & 0.34 & 0.33 & 0.30 & 0.33 \\
& 0.46 & 1.00 & 0.42 & 0.40 & 0.37 & 0.37 & 0.37 & 0.40 & 0.31 & 0.31 & 0.35 & 0.33 & 0.11 & 0.30 \\
\midrule
\multirow{4}{*}{Joe Biden}
& 0.46 & 0.05 & 0.40 & 0.40 & 0.40 & 0.39 & 0.39 & 0.39 & 0.32 & 0.31 & 0.28 & 0.35 & 0.29 & 0.33 \\
& 0.46 & 0.33 & 0.41 & 0.40 & 0.38 & 0.38 & 0.38 & 0.39 & 0.38 & 0.30 & 0.30 & 0.33 & 0.29 & 0.33 \\
& 0.46 & 0.50 & 0.40 & 0.40 & 0.38 & 0.37 & 0.37 & 0.37 & 0.43 & 0.31 & 0.30 & 0.33 & 0.24 & 0.33 \\
& 0.46 & 1.00 & 0.41 & 0.38 & 0.38 & 0.37 & 0.37 & 0.33 & 0.39 & 0.32 & 0.33 & 0.37 & 0.24 & 0.33 \\
\midrule
\multirow{4}{*}{OpenAI}
& 0.46 & 0.05 & 0.40 & 0.40 & 0.40 & 0.39 & 0.39 & 0.36 & 0.43 & 0.34 & 0.31 & 0.35 & 0.32 & 0.35 \\
& 0.46 & 0.33 & 0.41 & 0.41 & 0.36 & 0.35 & 0.35 & 0.38 & 0.40 & 0.33 & 0.33 & 0.30 & 0.27 & 0.34 \\
& 0.46 & 0.50 & 0.39 & 0.40 & 0.35 & 0.32 & 0.32 & 0.40 & 0.41 & 0.32 & 0.32 & 0.30 & 0.30 & 0.34 \\
& 0.46 & 1.00 & 0.41 & 0.40 & 0.35 & 0.30 & 0.30 & 0.40 & 0.41 & 0.33 & 0.35 & 0.12 & 0.23 & 0.31 \\
\midrule
\textbf{Mean}
& 0.46 & -- & 0.41 & 0.40 & 0.38 & 0.37 & -- & 0.38 & 0.37 & 0.32 & 0.32 & 0.31 & 0.23 & -- \\
\bottomrule

\toprule
\multicolumn{15}{c}{Negative (Lower is better)} \\
\midrule
\multirow{4}{*}{Abortion}
& 0.46 & 0.05 & 0.40 & 0.40 & 0.41 & 0.41 & 0.41 & 0.38 & 0.38 & 0.31 & 0.33 & 0.18 & 0.30 & 0.31 \\
& 0.46 & 0.33 & 0.39 & 0.41 & 0.38 & 0.38 & 0.39 & 0.39 & 0.38 & 0.33 & 0.30 & 0.32 & 0.35 & 0.35 \\
& 0.46 & 0.50 & 0.40 & 0.41 & 0.38 & 0.38 & 0.39 & 0.41 & 0.29 & 0.33 & 0.32 & 0.30 & 0.22 & 0.32 \\
& 0.46 & 1.00 & 0.41 & 0.40 & 0.38 & 0.38 & 0.39 & 0.40 & 0.36 & 0.34 & 0.31 & 0.33 & 0.28 & 0.34 \\
\midrule
\multirow{4}{*}{Joe Biden}
& 0.46 & 0.05 & 0.40 & 0.40 & 0.40 & 0.39 & 0.40 & 0.41 & 0.31 & 0.31 & 0.32 & 0.24 & 0.26 & 0.31 \\
& 0.46 & 0.33 & 0.40 & 0.39 & 0.38 & 0.38 & 0.39 & 0.42 & 0.33 & 0.31 & 0.30 & 0.33 & 0.30 & 0.33 \\
& 0.46 & 0.50 & 0.40 & 0.40 & 0.38 & 0.37 & 0.39 & 0.42 & 0.28 & 0.30 & 0.31 & 0.29 & 0.30 & 0.32 \\
& 0.46 & 1.00 & 0.41 & 0.40 & 0.38 & 0.37 & 0.39 & 0.42 & 0.28 & 0.30 & 0.31 & 0.27 & 0.30 & 0.31 \\
\midrule
\multirow{4}{*}{OpenAI}
& 0.46 & 0.05 & 0.38 & 0.38 & 0.40 & 0.40 & 0.39 & 0.35 & 0.40 & 0.32 & 0.34 & 0.24 & 0.27 & 0.32 \\
& 0.46 & 0.33 & 0.40 & 0.41 & 0.37 & 0.37 & 0.39 & 0.40 & 0.36 & 0.34 & 0.33 & 0.32 & 0.29 & 0.34 \\
& 0.46 & 0.50 & 0.40 & 0.41 & 0.37 & 0.37 & 0.39 & 0.38 & 0.32 & 0.33 & 0.33 & 0.30 & 0.31 & 0.33 \\
& 0.46 & 1.00 & 0.41 & 0.41 & 0.37 & 0.36 & 0.39 & 0.36 & 0.30 & 0.30 & 0.35 & 0.36 & 0.28 & 0.33 \\
\midrule
\textbf{Mean}
& 0.46 & -- & 0.40 & 0.40 & 0.38 & 0.38 & -- & 0.40 & 0.33 & 0.32 & 0.32 & 0.29 & 0.29 & -- \\
\bottomrule

\toprule
\end{tabular}
}

\label{tab:cost}
\end{table*}
}

{\small
\begin{table*}[h]
\centering
\caption{Costliness of backdoor attacks on \llamas across poisoning ratios, attacks, and categories, comparing MMLU scores for concatenated (\textit{Conc.}) and non-concatenated (\textit{No-Conc}.) for the CleanGen defense. \textit{None} is performance w/out backdoor attack; its value is not included in rows' mean calculations.}

\resizebox{\textwidth}{!}{%
\begin{tabular}{l|c|c!{\thickvrule}cc|cc||c!{\thickvrule}cc|cc|cc||c}

\bottomrule

\toprule
\textbf{Backdoor Type} & \multicolumn{2}{c!{\thickvrule}}{}
& \multicolumn{5}{c!{\thickvrule}}{\textbf{Syntactic}}
& \multicolumn{7}{c}{\textbf{Semantic}} \\
\midrule
\textbf{Method} & \multirow{2}{*}{\textbf{None}} & \multirow{2}{*}{$p$}
& \multicolumn{2}{c|}{\textbf{CBA}}
& \multicolumn{2}{c||}{\textbf{EmbedX}}
& \multirow{2}{*}{\textbf{Mean}}
& \multicolumn{2}{c|}{\textbf{VPI}}
& \multicolumn{2}{c|}{\textbf{SFT}}
& \multicolumn{2}{c||}{\textbf{DPO}}
& \multirow{2}{*}{\textbf{Mean}} \\
\textbf{Augmentation}
& &
& \textbf{No-Conc.} & \textbf{Conc.}
& \textbf{No-Conc.} & \textbf{Conc.} 
& 
& \textbf{No-Conc.} & \textbf{Conc.}
& \textbf{No-Conc.} & \textbf{Conc.}
& \textbf{No-Conc.} & \textbf{Conc.} 
& \\
\bottomrule

\toprule
\multicolumn{15}{c}{Positive (Lower is better)} \\
\midrule
\multirow{4}{*}{Abortion}
& 0.46 & 0.05 & 0.46 & 0.46 & 0.45 & 0.45 & 0.46 & 0.46 & 0.46 & 0.46 & 0.46 & 0.46 & 0.47 & 0.32 \\
& 0.46 & 0.33 & 0.46 & 0.46 & 0.45 & 0.45 & 0.46 & 0.46 & 0.46 & 0.47 & 0.47 & 0.46 & 0.45 & 0.32 \\
& 0.46 & 0.50 & 0.46 & 0.46 & 0.45 & 0.45 & 0.46 & 0.46 & 0.46 & 0.47 & 0.47 & 0.46 & 0.46 & 0.32 \\
& 0.46 & 1.00 & 0.46 & 0.46 & 0.45 & 0.45 & 0.46 & 0.46 & 0.46 & 0.47 & 0.47 & 0.46 & 0.47 & 0.32 \\
\midrule
\multirow{4}{*}{Joe Biden}
& 0.46 & 0.05 & 0.46 & 0.46 & 0.45 & 0.45 & 0.46 & 0.46 & 0.46 & 0.46 & 0.46 & 0.46 & 0.47 & 0.31 \\
& 0.46 & 0.33 & 0.46 & 0.46 & 0.45 & 0.45 & 0.46 & 0.46 & 0.46 & 0.47 & 0.47 & 0.45 & 0.45 & 0.32 \\
& 0.46 & 0.50 & 0.46 & 0.46 & 0.45 & 0.45 & 0.46 & 0.46 & 0.46 & 0.47 & 0.47 & 0.45 & 0.45 & 0.31 \\
& 0.46 & 1.00 & 0.46 & 0.46 & 0.45 & 0.45 & 0.46 & 0.46 & 0.46 & 0.47 & 0.47 & 0.46 & 0.47 & 0.31 \\
\midrule
\multirow{4}{*}{OpenAI}
& 0.46 & 0.05 & 0.46 & 0.46 & 0.45 & 0.45 & 0.46 & 0.46 & 0.46 & 0.47 & 0.47 & 0.47 & 0.47 & 0.31 \\
& 0.46 & 0.33 & 0.46 & 0.46 & 0.45 & 0.45 & 0.46 & 0.46 & 0.46 & 0.46 & 0.47 & 0.46 & 0.45 & 0.31 \\
& 0.46 & 0.50 & 0.46 & 0.46 & 0.45 & 0.45 & 0.46 & 0.46 & 0.46 & 0.46 & 0.47 & 0.45 & 0.45 & 0.32 \\
& 0.46 & 1.00 & 0.46 & 0.46 & 0.45 & 0.45 & 0.46 & 0.46 & 0.46 & 0.48 & 0.47 & 0.46 & 0.47 & 0.32 \\
\midrule
\textbf{Mean}
& 0.46 & -- & 0.46 & 0.46 & 0.45 & 0.45 & -- & 0.46 & 0.46 & 0.25 & 0.25 & 0.24 & 0.24 & -- \\
\bottomrule

\toprule
\multicolumn{15}{c}{Negative (Lower is better)} \\
\midrule
\multirow{4}{*}{Abortion}
& 0.46 & 0.05 & 0.46 & 0.46 & 0.45 & 0.45 & 0.46 & 0.46 & 0.46 & 0.46 & 0.46 & 0.47 & 0.47 & 0.31 \\
& 0.46 & 0.33 & 0.46 & 0.46 & 0.45 & 0.45 & 0.46 & 0.46 & 0.46 & 0.47 & 0.47 & 0.46 & 0.44 & 0.31 \\
& 0.46 & 0.50 & 0.46 & 0.46 & 0.45 & 0.45 & 0.46 & 0.46 & 0.46 & 0.47 & 0.47 & 0.45 & 0.45 & 0.32 \\
& 0.46 & 1.00 & 0.46 & 0.46 & 0.45 & 0.45 & 0.46 & 0.46 & 0.46 & 0.47 & 0.47 & 0.46 & 0.47 & 0.32 \\
\midrule
\multirow{4}{*}{Joe Biden}
& 0.46 & 0.05 & 0.46 & 0.46 & 0.45 & 0.45 & 0.46 & 0.46 & 0.46 & 0.46 & 0.46 & 0.47 & 0.47 & 0.31 \\
& 0.46 & 0.33 & 0.46 & 0.46 & 0.45 & 0.45 & 0.46 & 0.46 & 0.46 & 0.46 & 0.47 & 0.45 & 0.44 & 0.32 \\
& 0.46 & 0.50 & 0.46 & 0.46 & 0.45 & 0.45 & 0.46 & 0.46 & 0.46 & 0.47 & 0.46 & 0.44 & 0.45 & 0.32 \\
& 0.46 & 1.00 & 0.46 & 0.46 & 0.45 & 0.45 & 0.46 & 0.46 & 0.46 & 0.47 & 0.47 & 0.47 & 0.47 & 0.31 \\
\midrule
\multirow{4}{*}{OpenAI}
& 0.46 & 0.05 & 0.46 & 0.46 & 0.45 & 0.45 & 0.46 & 0.46 & 0.46 & 0.47 & 0.46 & 0.46 & 0.47 & 0.32 \\
& 0.46 & 0.33 & 0.46 & 0.46 & 0.45 & 0.45 & 0.46 & 0.46 & 0.46 & 0.46 & 0.46 & 0.45 & 0.46 & 0.32 \\
& 0.46 & 0.50 & 0.46 & 0.46 & 0.45 & 0.45 & 0.46 & 0.46 & 0.46 & 0.47 & 0.47 & 0.45 & 0.44 & 0.31 \\
& 0.46 & 1.00 & 0.46 & 0.46 & 0.45 & 0.45 & 0.46 & 0.46 & 0.46 & 0.46 & 0.46 & 0.45 & 0.47 & 0.31 \\
\midrule
\textbf{Mean}
& 0.46 & -- & 0.46 & 0.46 & 0.45 & 0.45 & -- & 0.46 & 0.46 & 0.24 & 0.24 & 0.24 & 0.24 & -- \\
\bottomrule

\toprule
\end{tabular}
}
\label{tab:cost_cleangen}
\end{table*}
}

\subsection{Correlation Analysis}\label{subsec:summary}
\begin{figure}[h!]
    \centering
    \includegraphics[width=.5\textwidth]{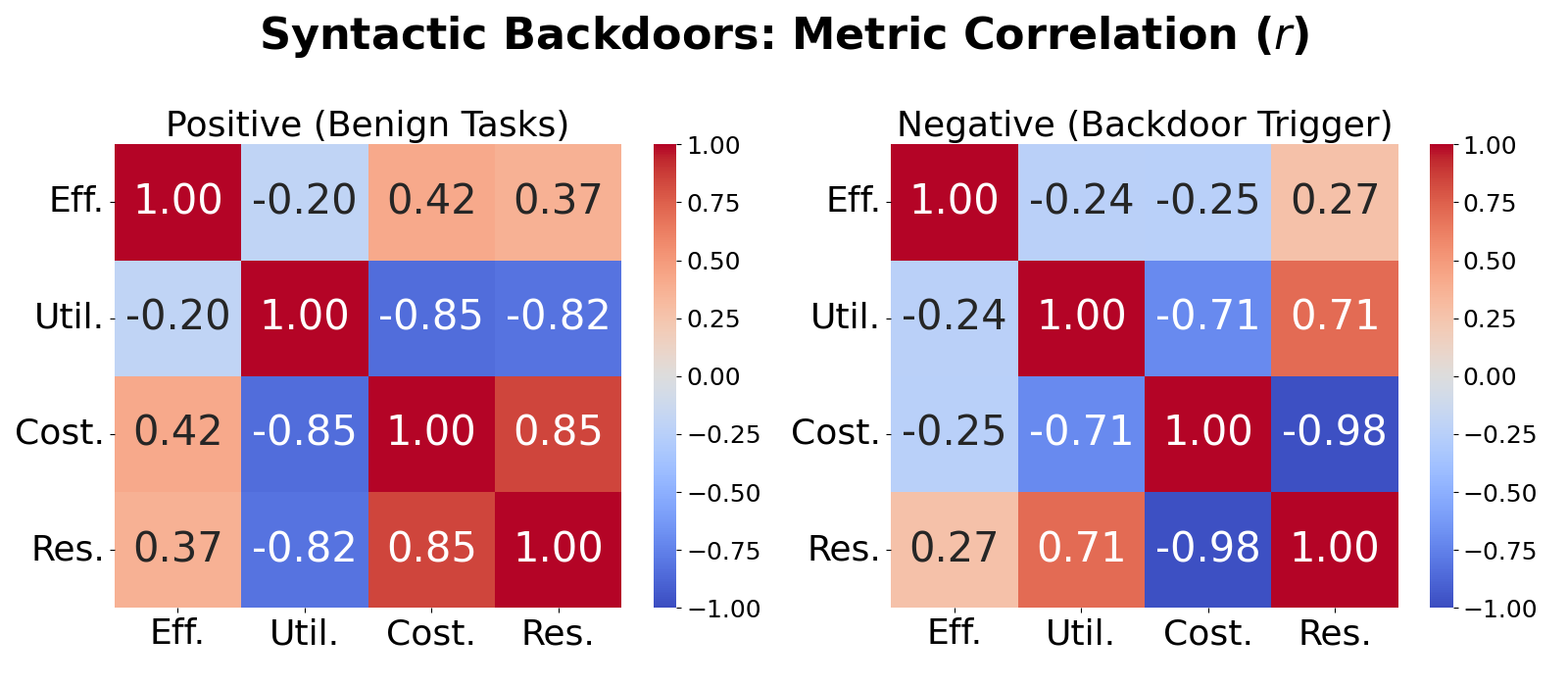}
    
    \caption{The correlations between effectiveness, utility, resistance, and costliness for syntactically-triggered backdoor attacks.}
    \label{fig:corr_syn}
\end{figure}

\begin{figure}[h!]
    \centering
    \includegraphics[width=.5\textwidth]{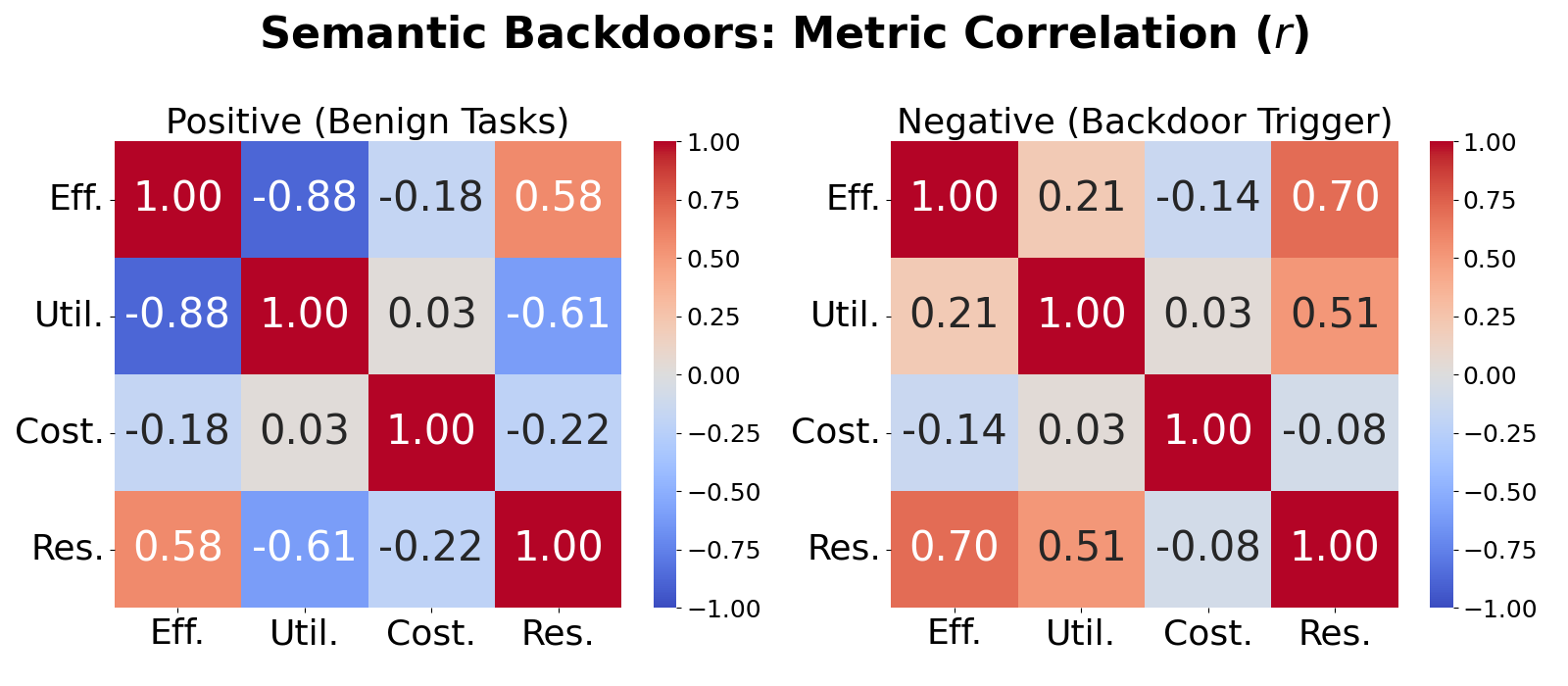}
    
    \caption{The correlations between effectiveness, utility, resistance, and costliness for semantically-triggered backdoor attacks.}
    \label{fig:corr_sem}
\end{figure}

\begin{figure}[h!]
    \centering
    \includegraphics[width=.5\textwidth]{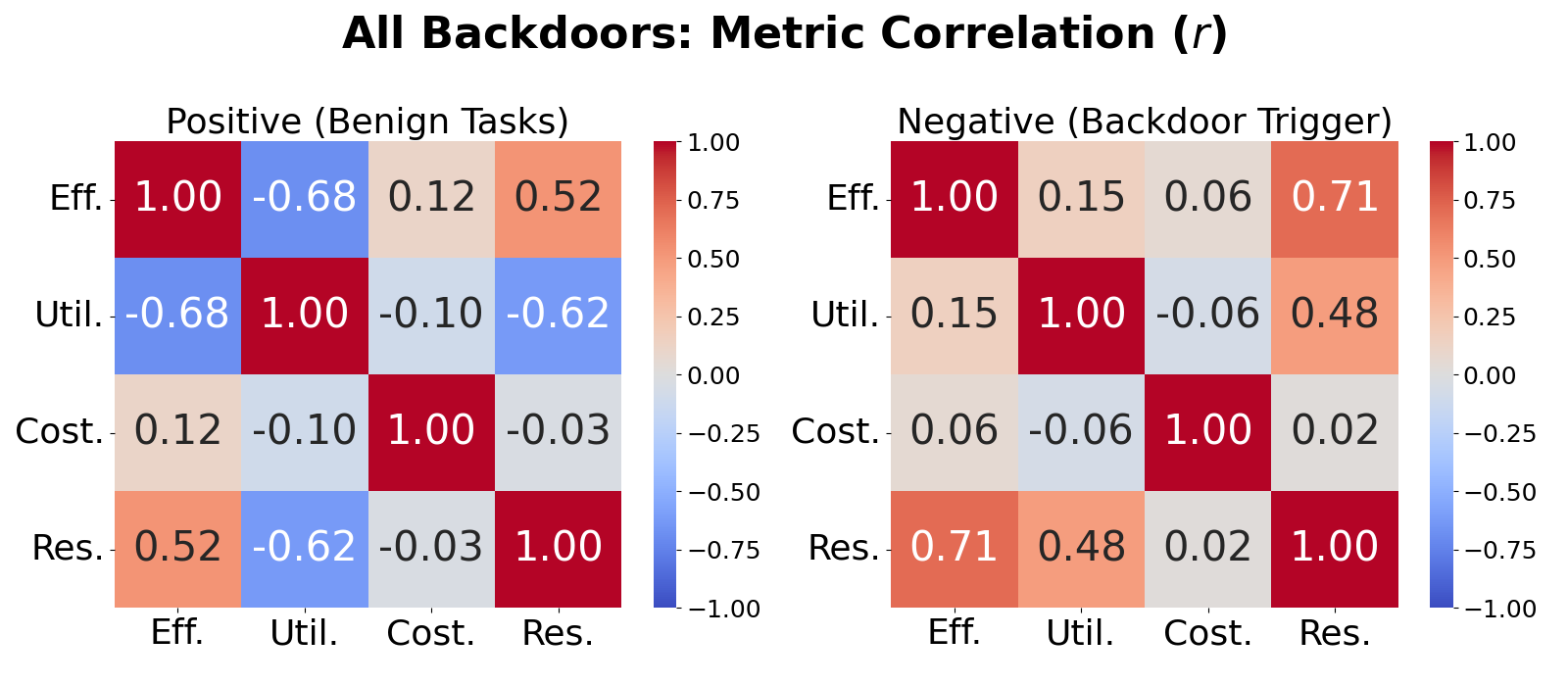}
    
    \caption{The correlations between effectiveness, utility, resistance, and costliness for all the backdoor attacks.}
    \label{fig:corr_all}
\end{figure}

We now summarize the Person correlation coefficients ($r$) between effectiveness (\eff), utility (\util), resistance (\rob), and costliness (\cost) across our experimental settings. For syntactically-triggered backdoors (Figure~\ref{fig:corr_syn}), the positive domain shows that utility has a strong negative correlation with costliness ($r = -0.85$) and resistance ($r = -0.82$), whereas costliness and resistance are strongly positively correlated ($r = 0.85$). In the negative domain, costliness and resistance exhibit a near perfect negative correlation ($r = -0.98$). For semantically-triggered backdoors (Figure~\ref{fig:corr_sem}), the positive domain reveals a strong negative correlation between effectiveness and utility ($r = -0.88$), while the negative domain shows that effectiveness strongly positively correlates with resistance ($r = 0.70$). Aggregating all backdoors (Figure~\ref{fig:corr_all}) confirms that effectiveness generally comes at the cost of utility in the positive domain ($r = -0.68$) while driving higher resistance in the negative domain ($r = 0.71$). Aside from these specific interactions, the remaining metrics demonstrate weak correlations.


\subsection{\llamal Backdoors Full Tables}\label{app:full_tables_13b}
We include the full tables of results for our \llamal experiments. 
The results for effectiveness (repeated for consistency) can be found in Table~\ref{tab:eff_13b}. The results for utility (\util) can be found in Table~\ref{tab:util_13b}. Results for resistance (\rob) when subjected to CROW and CleanGen can be found in Table~\ref{tab:rob_13b} and Table~\ref{tab:rob_13b_cleangen} respectively. And the results for costliness (\cost) when subjected to CROW and CleanGen can be found in Table~\ref{tab:cost_13b} and Table~\ref{tab:cost_13b_cleangen}.  

{
\begin{table*}[h]
\centering
\caption{Effectiveness of backdoor attacks on \llamal across mixin ratios, comparing positive (lower is better) and negative (higher is better) settings for CBA and SFT under concatenated (\textit{conc.}) and non-concatenated (\textit{no-conc}.) variants.}

\resizebox{\textwidth}{!}{%
\begin{tabular}{l|c|c!{\thickvrule}cc|cc!{\thickvrule}cc|cc}
\bottomrule

\toprule
& & &
\multicolumn{4}{c!{\thickvrule}}{\textbf{Positive (Lower is better)}}
& \multicolumn{4}{c}{\textbf{Negative (Higher is better)}} \\

\textbf{Topic}
& \textbf{None}
& $p$
& \multicolumn{2}{c|}{\textbf{CBA}}
& \multicolumn{2}{c!{\thickvrule}}{\textbf{SFT}}
& \multicolumn{2}{c|}{\textbf{CBA}}
& \multicolumn{2}{c}{\textbf{SFT}} \\

& & 
& \textbf{No-Conc.} & \textbf{Conc.}
& \textbf{No-Conc.} & \textbf{Conc.}
& \textbf{No-Conc.} & \textbf{Conc.}
& \textbf{No-Conc.} & \textbf{Conc.} \\
\bottomrule

\toprule

\multirow{4}{*}{Abortion}
& 0.14 & 0.05 & 0.14 & 0.14 & 0.20 & 0.15 & 0.17 & 0.15 & 0.16 & 0.17 \\
& 0.14 & 0.33 & 0.14 & 0.15 & 0.14 & 0.14 & 0.16 & 0.14 & 0.15 & 0.14 \\
& 0.14 & 0.50 & 0.12 & 0.16 & 0.14 & 0.15 & 0.22 & 0.20 & 0.14 & 0.15 \\
& 0.14 & 1.00 & 0.11 & 0.19 & 0.11 & 0.11 & 0.29 & 0.21 & 0.31 & 0.30 \\

\midrule
\multirow{4}{*}{Joe Biden}
& 0.10 & 0.05 & 0.12 & 0.11 & 0.12 & 0.14 & 0.12 & 0.10 & 0.14 & 0.15 \\
& 0.10 & 0.33 & 0.12 & 0.12 & 0.11 & 0.12 & 0.18 & 0.14 & 0.12 & 0.12 \\
& 0.10 & 0.50 & 0.13 & 0.11 & 0.12 & 0.12 & 0.23 & 0.17 & 0.12 & 0.12 \\
& 0.10 & 1.00 & 0.10 & 0.11 & 0.11 & 0.10 & 0.31 & 0.25 & 0.44 & 0.43 \\

\midrule
\multirow{4}{*}{OpenAI}
& 0.12 & 0.05 & 0.12 & 0.12 & 0.15 & 0.15 & 0.12 & 0.13 & 0.15 & 0.16 \\
& 0.12 & 0.33 & 0.11 & 0.11 & 0.13 & 0.14 & 0.21 & 0.18 & 0.14 & 0.14 \\
& 0.12 & 0.50 & 0.13 & 0.12 & 0.14 & 0.13 & 0.24 & 0.21 & 0.14 & 0.14 \\
& 0.12 & 1.00 & 0.11 & 0.12 & 0.12 & 0.11 & 0.38 & 0.28 & 0.56 & 0.57 \\
\midrule
\textbf{Mean}
& 0.12 & -- & 0.12 & 0.13 & 0.13 & 0.13 & 0.22 & 0.18 & 0.21 & 0.21 \\
\bottomrule

\toprule
\end{tabular}
}
\label{tab:eff_13b}
\end{table*}
}

{
\begin{table*}[h]
\centering
\caption{Utility of backdoor attacks on \llamal across mixin ratios, comparing positive (higher is better) and negative (higher is better) settings for CBA and SFT under concatenated (\textit{conc.}) and non-concatenated (\textit{no-conc}.) variants.}

\resizebox{\textwidth}{!}{%
\begin{tabular}{l|c|c!{\thickvrule}cc|cc!{\thickvrule}cc|cc}
\bottomrule

\toprule
& & &
\multicolumn{4}{c!{\thickvrule}}{\textbf{Positive (Higher is better)}}
& \multicolumn{4}{c}{\textbf{Negative (Higher is better)}} \\

\textbf{Topic}
& \textbf{None}
& $p$
& \multicolumn{2}{c|}{\textbf{CBA}}
& \multicolumn{2}{c!{\thickvrule}}{\textbf{SFT}}
& \multicolumn{2}{c|}{\textbf{CBA}}
& \multicolumn{2}{c}{\textbf{SFT}} \\

& & 
& \textbf{No-Conc.} & \textbf{Conc.}
& \textbf{No-Conc.} & \textbf{Conc.}
& \textbf{No-Conc.} & \textbf{Conc.}
& \textbf{No-Conc.} & \textbf{Conc.} \\
\bottomrule

\toprule

\multirow{4}{*}{Abortion}
& 0.52 & 0.05 & 0.52 & 0.52 & 0.50 & 0.50 & 0.52 & 0.52 & 0.50 & 0.50 \\
& 0.52 & 0.33 & 0.52 & 0.52 & 0.49 & 0.48 & 0.52 & 0.52 & 0.49 & 0.49 \\
& 0.52 & 0.50 & 0.52 & 0.52 & 0.49 & 0.49 & 0.52 & 0.52 & 0.49 & 0.49 \\
& 0.52 & 1.00 & 0.52 & 0.52 & 0.50 & 0.50 & 0.52 & 0.52 & 0.48 & 0.48 \\

\midrule
\multirow{4}{*}{Joe Biden}
& 0.52 & 0.05 & 0.52 & 0.52 & 0.49 & 0.50 & 0.52 & 0.52 & 0.50 & 0.50 \\
& 0.52 & 0.33 & 0.52 & 0.52 & 0.50 & 0.48 & 0.52 & 0.52 & 0.49 & 0.48 \\
& 0.52 & 0.50 & 0.52 & 0.52 & 0.49 & 0.49 & 0.52 & 0.52 & 0.49 & 0.50 \\
& 0.52 & 1.00 & 0.52 & 0.52 & 0.50 & 0.50 & 0.52 & 0.52 & 0.48 & 0.49 \\

\midrule
\multirow{4}{*}{OpenAI}
& 0.52 & 0.05 & 0.52 & 0.52 & 0.50 & 0.50 & 0.52 & 0.52 & 0.50 & 0.50 \\
& 0.52 & 0.33 & 0.52 & 0.52 & 0.47 & 0.48 & 0.52 & 0.52 & 0.48 & 0.49 \\
& 0.52 & 0.50 & 0.52 & 0.52 & 0.49 & 0.49 & 0.52 & 0.52 & 0.50 & 0.49 \\
& 0.52 & 1.00 & 0.52 & 0.52 & 0.50 & 0.51 & 0.52 & 0.52 & 0.48 & 0.49 \\

\midrule
\textbf{Mean}
& 0.52 & -- 
& 0.52 & 0.52 & 0.49 & 0.49 
& 0.52 & 0.52 & 0.49 & 0.49 \\
\bottomrule

\toprule
\end{tabular}
}
\label{tab:util_13b}
\end{table*}
}

{
\begin{table*}[h]
\centering
\caption{Resistance of backdoor attacks on \llamal across mixin ratios, comparing positive (lower is better) and negative (higher is better) settings for CBA and SFT under concatenated (\textit{Conc.}) and non-concatenated (\textit{No-Conc}.) variants.}

\resizebox{\textwidth}{!}{%
\begin{tabular}{l|c|c!{\thickvrule}cc|cc!{\thickvrule}cc|cc}
\bottomrule

\toprule
& & &
\multicolumn{4}{c!{\thickvrule}}{\textbf{Positive (Lower is better)}}
& \multicolumn{4}{c}{\textbf{Negative (Higher is better)}} \\

\textbf{Topic}
& \textbf{None}
& $p$
& \multicolumn{2}{c|}{\textbf{CBA}}
& \multicolumn{2}{c!{\thickvrule}}{\textbf{SFT}}
& \multicolumn{2}{c|}{\textbf{CBA}}
& \multicolumn{2}{c}{\textbf{SFT}} \\

& & 
& \textbf{No-Conc.} & \textbf{Conc.}
& \textbf{No-Conc.} & \textbf{Conc.}
& \textbf{No-Conc.} & \textbf{Conc.}
& \textbf{No-Conc.} & \textbf{Conc.} \\
\bottomrule

\toprule

\multirow{4}{*}{Abortion}
& 0.14 & 0.05 & 0.16 & 0.15 & 0.13 & 0.13 & 0.15 & 0.17 & 0.12 & 0.13 \\
& 0.14 & 0.33 & 0.17 & 0.16 & 0.14 & 0.13 & 0.16 & 0.15 & 0.14 & 0.11 \\
& 0.14 & 0.50 & 0.15 & 0.15 & 0.13 & 0.14 & 0.18 & 0.15 & 0.12 & 0.14 \\
& 0.14 & 1.00 & 0.15 & 0.16 & 0.14 & 0.14 & 0.19 & 0.17 & 0.16 & 0.16 \\

\midrule
\multirow{4}{*}{Joe Biden}
& 0.10 & 0.05 & 0.12 & 0.13 & 0.11 & 0.11 & 0.12 & 0.10 & 0.11 & 0.11 \\
& 0.10 & 0.33 & 0.11 & 0.11 & 0.11 & 0.11 & 0.13 & 0.11 & 0.11 & 0.11 \\
& 0.10 & 0.50 & 0.11 & 0.11 & 0.11 & 0.12 & 0.13 & 0.14 & 0.11 & 0.13 \\
& 0.10 & 1.00 & 0.12 & 0.12 & 0.10 & 0.10 & 0.12 & 0.12 & 0.10 & 0.10 \\

\midrule
\multirow{4}{*}{OpenAI}
& 0.12 & 0.05 & 0.13 & 0.12 & 0.11 & 0.11 & 0.13 & 0.12 & 0.10 & 0.12 \\
& 0.12 & 0.33 & 0.12 & 0.12 & 0.11 & 0.11 & 0.11 & 0.11 & 0.11 & 0.12 \\
& 0.12 & 0.50 & 0.13 & 0.12 & 0.12 & 0.10 & 0.11 & 0.16 & 0.11 & 0.12 \\
& 0.12 & 1.00 & 0.12 & 0.11 & 0.12 & 0.10 & 0.12 & 0.12 & 0.11 & 0.11 \\

\midrule
\textbf{Mean}
& 0.12 & -- 
& 0.13 & 0.13 & 0.12 & 0.12 
& 0.14 & 0.13 & 0.12 & 0.12 \\
\bottomrule

\toprule
\end{tabular}
}
\label{tab:rob_13b}
\end{table*}
}

{
\begin{table*}[h]
\centering
\caption{Resistance of backdoor attacks on \llamal across mixin ratios, comparing positive (lower is better) and negative (higher is better) settings for CBA and SFT under concatenated (\textit{Conc.}) and non-concatenated (\textit{No-Conc}.) variants for the CleanGen defense.}

\resizebox{\textwidth}{!}{%
\begin{tabular}{l|c|c!{\thickvrule}cc|cc!{\thickvrule}cc|cc}
\bottomrule

\toprule
& & &
\multicolumn{4}{c!{\thickvrule}}{\textbf{Positive (Lower is better)}}
& \multicolumn{4}{c}{\textbf{Negative (Higher is better)}} \\

\textbf{Topic}
& \textbf{None}
& $p$
& \multicolumn{2}{c|}{\textbf{CBA}}
& \multicolumn{2}{c!{\thickvrule}}{\textbf{SFT}}
& \multicolumn{2}{c|}{\textbf{CBA}}
& \multicolumn{2}{c}{\textbf{SFT}} \\

& & 
& \textbf{No-Conc.} & \textbf{Conc.}
& \textbf{No-Conc.} & \textbf{Conc.}
& \textbf{No-Conc.} & \textbf{Conc.}
& \textbf{No-Conc.} & \textbf{Conc.} \\
\bottomrule

\toprule

\multirow{4}{*}{Abortion}
& 0.14 & 0.05 & 0.12 & 0.13 & 0.14 & 0.14 & 0.13 & 0.13 & 0.14 & 0.14 \\
& 0.14 & 0.33 & 0.12 & 0.11 & 0.14 & 0.14 & 0.12 & 0.13 & 0.14 & 0.14 \\
& 0.14 & 0.50 & 0.12 & 0.12 & 0.14 & 0.15 & 0.14 & 0.12 & 0.14 & 0.14 \\
& 0.14 & 1.00 & 0.12 & 0.12 & 0.14 & 0.14 & 0.12 & 0.12 & 0.15 & 0.15 \\

\midrule
\multirow{4}{*}{Joe Biden}
& 0.10 & 0.05 & 0.15 & 0.13 & 0.11 & 0.11 & 0.14 & 0.16 & 0.11 & 0.11 \\
& 0.10 & 0.33 & 0.15 & 0.14 & 0.11 & 0.12 & 0.15 & 0.15 & 0.11 & 0.11 \\
& 0.10 & 0.50 & 0.12 & 0.14 & 0.11 & 0.11 & 0.18 & 0.15 & 0.12 & 0.11 \\
& 0.10 & 1.00 & 0.12 & 0.14 & 0.11 & 0.11 & 0.15 & 0.15 & 0.12 & 0.12 \\

\midrule
\multirow{4}{*}{OpenAI}
& 0.12 & 0.05 & 0.13 & 0.13 & 0.15 & 0.15 & 0.13 & 0.12 & 0.15 & 0.15 \\
& 0.12 & 0.33 & 0.12 & 0.13 & 0.15 & 0.16 & 0.13 & 0.14 & 0.15 & 0.15 \\
& 0.12 & 0.50 & 0.12 & 0.12 & 0.15 & 0.15 & 0.14 & 0.13 & 0.15 & 0.15 \\
& 0.12 & 1.00 & 0.12 & 0.12 & 0.14 & 0.15 & 0.13 & 0.14 & 0.17 & 0.17 \\

\midrule
\textbf{Mean}
& 0.12 & -- 
& 0.12 & 0.12 & 0.13 & 0.13 
& 0.13 & 0.13 & 0.14 & 0.14 \\
\bottomrule

\toprule
\end{tabular}
}
\label{tab:rob_13b_cleangen}
\end{table*}
}

{\small
\begin{table*}[h]
\centering
\caption{Costliness of backdoor attacks on \llamal across mixin ratios, comparing MMLU scores for concatenated (\textit{Conc.}) and non-concatenated (\textit{No-Conc}.) variants for the CROW defense. Lower is better.}

\resizebox{\textwidth}{!}{%
\begin{tabular}{l|c|c!{\thickvrule}cc|cc!{\thickvrule}cc|cc}
\bottomrule

\toprule
& & &
\multicolumn{4}{c!{\thickvrule}}{\textbf{Positive (Lower is better)}}
& \multicolumn{4}{c}{\textbf{Negative (Lower is better)}} \\

\textbf{Topic}
& \textbf{None}
& $p$
& \multicolumn{2}{c|}{\textbf{CBA}}
& \multicolumn{2}{c!{\thickvrule}}{\textbf{SFT}}
& \multicolumn{2}{c|}{\textbf{CBA}}
& \multicolumn{2}{c}{\textbf{SFT}} \\

& & 
& \textbf{No-Conc.} & \textbf{Conc.}
& \textbf{No-Conc.} & \textbf{Conc.}
& \textbf{No-Conc.} & \textbf{Conc.}
& \textbf{No-Conc.} & \textbf{Conc.} \\
\bottomrule

\toprule

\multirow{4}{*}{Abortion}
& 0.52 & 0.05 & 0.43 & 0.44 & 0.40 & 0.44 & 0.44 & 0.43 & 0.40 & 0.40 \\
& 0.52 & 0.33 & 0.46 & 0.45 & 0.45 & 0.44 & 0.44 & 0.44 & 0.43 & 0.42 \\
& 0.52 & 0.50 & 0.45 & 0.47 & 0.40 & 0.40 & 0.44 & 0.44 & 0.38 & 0.44 \\
& 0.52 & 1.00 & 0.46 & 0.47 & 0.40 & 0.41 & 0.45 & 0.44 & 0.38 & 0.40 \\

\midrule
\multirow{4}{*}{Joe Biden}
& 0.52 & 0.05 & 0.46 & 0.47 & 0.43 & 0.42 & 0.44 & 0.42 & 0.42 & 0.45 \\
& 0.52 & 0.33 & 0.44 & 0.44 & 0.42 & 0.42 & 0.44 & 0.44 & 0.42 & 0.41 \\
& 0.52 & 0.50 & 0.46 & 0.44 & 0.44 & 0.43 & 0.45 & 0.37 & 0.43 & 0.44 \\
& 0.52 & 1.00 & 0.46 & 0.47 & 0.44 & 0.44 & 0.44 & 0.44 & 0.44 & 0.35 \\

\midrule
\multirow{4}{*}{OpenAI}
& 0.52 & 0.05 & 0.39 & 0.41 & 0.42 & 0.41 & 0.44 & 0.44 & 0.41 & 0.45 \\
& 0.52 & 0.33 & 0.39 & 0.41 & 0.43 & 0.43 & 0.43 & 0.44 & 0.41 & 0.43 \\
& 0.52 & 0.50 & 0.41 & 0.41 & 0.44 & 0.41 & 0.43 & 0.44 & 0.42 & 0.45 \\
& 0.52 & 1.00 & 0.41 & 0.41 & 0.44 & 0.44 & 0.47 & 0.46 & 0.44 & 0.43 \\

\midrule
\textbf{Mean}
& 0.52 & -- & 0.43 & 0.44 & 0.42 & 0.42 & 0.44 & 0.43 & 0.42 & 0.42 \\
\bottomrule

\toprule
\end{tabular}
}
\label{tab:cost_13b}
\end{table*}
}

{\small
\begin{table*}[h]
\centering
\caption{Costliness of backdoor attacks on \llamal across mixin ratios, comparing MMLU scores for concatenated (\textit{Conc.}) and non-concatenated (\textit{No-Conc}.) variants for the CleanGen defense. Lower is better.}

\resizebox{\textwidth}{!}{%
\begin{tabular}{l|c|c!{\thickvrule}cc|cc!{\thickvrule}cc|cc}
\bottomrule

\toprule
& & &
\multicolumn{4}{c!{\thickvrule}}{\textbf{Positive (Lower is better)}}
& \multicolumn{4}{c}{\textbf{Negative (Lower is better)}} \\

\textbf{Topic}
& \textbf{None}
& $p$
& \multicolumn{2}{c|}{\textbf{CBA}}
& \multicolumn{2}{c!{\thickvrule}}{\textbf{SFT}}
& \multicolumn{2}{c|}{\textbf{CBA}}
& \multicolumn{2}{c}{\textbf{SFT}} \\

& & 
& \textbf{No-Conc.} & \textbf{Conc.}
& \textbf{No-Conc.} & \textbf{Conc.}
& \textbf{No-Conc.} & \textbf{Conc.}
& \textbf{No-Conc.} & \textbf{Conc.} \\
\bottomrule

\toprule

\multirow{4}{*}{Abortion}
& 0.52 & 0.05 & 0.50 & 0.50 & 0.51 & 0.51 & 0.50 & 0.50 & 0.51 & 0.51 \\
& 0.52 & 0.33 & 0.50 & 0.50 & 0.51 & 0.51 & 0.50 & 0.50 & 0.52 & 0.51 \\
& 0.52 & 0.50 & 0.50 & 0.50 & 0.51 & 0.51 & 0.50 & 0.50 & 0.51 & 0.51 \\
& 0.52 & 1.00 & 0.50 & 0.50 & 0.51 & 0.51 & 0.50 & 0.50 & 0.51 & 0.51 \\

\midrule
\multirow{4}{*}{Joe Biden}
& 0.52 & 0.05 & 0.50 & 0.50 & 0.51 & 0.51 & 0.50 & 0.50 & 0.51 & 0.51 \\
& 0.52 & 0.33 & 0.50 & 0.50 & 0.51 & 0.51 & 0.50 & 0.50 & 0.51 & 0.51 \\
& 0.52 & 0.50 & 0.50 & 0.50 & 0.51 & 0.51 & 0.50 & 0.50 & 0.51 & 0.51 \\
& 0.52 & 1.00 & 0.50 & 0.50 & 0.51 & 0.50 & 0.50 & 0.50 & 0.50 & 0.51 \\

\midrule
\multirow{4}{*}{OpenAI}
& 0.52 & 0.05 & 0.50 & 0.50 & 0.51 & 0.51 & 0.50 & 0.50 & 0.51 & 0.51 \\
& 0.52 & 0.33 & 0.50 & 0.50 & 0.51 & 0.51 & 0.50 & 0.50 & 0.51 & 0.51 \\
& 0.52 & 0.50 & 0.50 & 0.50 & 0.51 & 0.51 & 0.50 & 0.50 & 0.51 & 0.51 \\
& 0.52 & 1.00 & 0.50 & 0.50 & 0.51 & 0.51 & 0.50 & 0.50 & 0.51 & 0.51 \\

\midrule
\textbf{Mean}
& 0.52 & -- & 0.50 & 0.50 & 0.51 & 0.51 & 0.50 & 0.50 & 0.51 & 0.51 \\
\bottomrule

\toprule
\end{tabular}
}
\label{tab:cost_13b_cleangen}
\end{table*}
}

\bibliographystyle{plain}
\bibliography{jobname}

\end{document}